\RequirePackage[l2tabu]{nag}		
\documentclass[a4paper,11pt,leqno,openbib]{memoir} 
\usepackage{datetime}
\usepackage{ifpdf}
\ifpdf
\fi
\ifdraftdoc 
	\usepackage{draftwatermark}				
	\SetWatermarkScale{0.3}
	\SetWatermarkText{\bf Draft: \today}
\fi
\newsubfloat{figure}
\newsubfloat{table}
\settrimmedsize{297mm}{210mm}{*}
\settypeblocksize{634pt}{448.13pt}{*} 
\setulmargins{4cm}{*}{*} 
\setlrmargins{*}{*}{1.5} 
\setmarginnotes{17pt}{51pt}{\onelineskip} 
\setheadfoot{\onelineskip}{2\onelineskip} 
\setheaderspaces{*}{2\onelineskip}{*} 
\checkandfixthelayout
\usepackage{fouriernc}
\usepackage[T1]{fontenc}
\OnehalfSpacing 
\setsecnumdepth{subsection} 
\maxsecnumdepth{subsubsection}
\usepackage{calc,soul,fourier}
\makeatletter 
\newlength\dlf@normtxtw 
\newsavebox{\feline@chapter} 
\newcommand\feline@chapter@marker[1][4cm]{%
	\sbox\feline@chapter{%
		\resizebox{!}{#1}{\fboxsep=1pt%
			\colorbox{gray}{\color{white}\thechapter}%
		}}%
		\rotatebox{90}{%
			\resizebox{%
				\heightof{\usebox{\feline@chapter}}+\depthof{\usebox{\feline@chapter}}}%
			{!}{\scshape\so\@chapapp}}\quad%
		\raisebox{\depthof{\usebox{\feline@chapter}}}{\usebox{\feline@chapter}}%
} 
\newcommand\feline@chm[1][4cm]{%
	\sbox\feline@chapter{\feline@chapter@marker[#1]}%
	\makebox[0pt][c]{
		\makebox[1cm][r]{\usebox\feline@chapter}%
	}}
\makechapterstyle{daleifmodif}{

	\renewcommand\printchapternum{\null\hfill\feline@chm[2.5cm]\par}

} 
\makeatother 
\chapterstyle{daleifmodif}
\makepagestyle{myvf} 
\makeoddfoot{myvf}{}{\thepage}{} 
\makeevenfoot{myvf}{}{\thepage}{} 
\makeheadrule{myvf}{\textwidth}{\normalrulethickness} 
\makeevenhead{myvf}{\small\textsc{\leftmark}}{}{} 
\makeoddhead{myvf}{\small\textsc{\leftmark}}{}{}
\usepackage{color}

\newcommand{\bv}{{\mathbf v}}
\newcommand{\ba}{{\mathbf a}}
\newcommand{\br}{{\mathbf r}}
\newcommand{\bl}{{\mathbf l}}
\newcommand{\bc}{{\mathbf c}}

\makeindex
\usepackage{import}

\usepackage{lipsum}					
\usepackage{amsfonts} 					
\usepackage[centertags]{amsmath}			
\usepackage{stmaryrd}					
\usepackage{amssymb}					
\usepackage{amsthm}					
\usepackage{newlfont}					
\usepackage{graphicx}					
\usepackage{longtable,rotating}			
\usepackage[applemac]{inputenc}			
\usepackage{colortbl}					
\usepackage{wasysym}					
\usepackage{mathrsfs}					
\usepackage{float}						
\usepackage{verbatim}					
\usepackage{upgreek }					
\usepackage{latexsym}					
\usepackage[square,numbers,
		     sort&compress]{natbib}		
\usepackage{url}						
\usepackage{etex}						
\usepackage{fixltx2e}					
\usepackage[spanish,english]{babel}		
\usepackage{color}                    				
\usepackage[colorlinks=true,
		     allcolors=black]{hyperref}              
\usepackage{memhfixc}					
\usepackage{enumerate}					
\usepackage{footnote}					
\usepackage{microtype}					
\usepackage{rotfloat}					
\usepackage{alltt}						
\usepackage[version=0.96]{pgf}			
\usepackage{tikz}						
\usetikzlibrary{arrows,shapes,snakes,
		       automata,backgrounds,
		       petri,topaths}				

\usepackage{lscape}
\usepackage{enumitem}
\usepackage{caption}
\usepackage{pifont}
\usepackage{tikz}
\usetikzlibrary{matrix}

\subcaptionlabelfont{\scshape\small}
\subcaptionfont{\scshape\small}

\newcommand{\pgftextcircled}[1]{                                                                    
    \setbox0=\hbox{#1}%
    \dimen0\wd0%
    \divide\dimen0 by 2%
    \begin{tikzpicture}[baseline=(a.base)]%
        \useasboundingbox (-\the\dimen0,0pt) rectangle (\the\dimen0,1pt);
        \node[circle,draw,outer sep=0pt,inner sep=0.1ex] (a) {#1};
    \end{tikzpicture}
}
\newcommand{\blackged}{\hfill$\blacksquare$}
\newcommand{\whiteged}{\hfill$\square$}
\newcounter{proofcount}

\let\oldsqrt\sqrt
\def\sqrt{\mathpalette\DHLhksqrt}
\def\DHLhksqrt#1#2{%
\setbox0=\hbox{$#1\oldsqrt{#2\,}$}\dimen0=\ht0
\advance\dimen0-0.2\ht0
\setbox2=\hbox{\vrule height\ht0 depth -\dimen0}%
{\box0\lower0.4pt\box2}}
\newcommand{\mycaption}[2][\@empty]{
	\captionnamefont{\scshape \bfseries}
    \captiontitlefont{\scshape}
	\changecaptionwidth
	\captionwidth{0.9\linewidth}
	\captiondelim{:\:} 
	\indentcaption{0.75cm}
	\captionstyle[\centering]{}
	\setlength{\belowcaptionskip}{10pt}
	\ifx \@empty#1 \caption{#2}\else \caption[#1]{#2}
}
\usepackage{lettrine}
\newcommand{\initial}[1]{%
	\lettrine[lines=3,lhang=0.33,nindent=0em]{
		\color{gray}
     		{\textsc{#1}}}{}}
\theoremstyle{plain}

\theoremstyle{plain}

\theoremstyle{plain}
\theoremstyle{definition}

\theoremstyle{plain}

\theoremstyle{plain}

\theoremstyle{plain}

\begin{document}

\let\oldcleartorecto\cleartorecto 
\let\cleartorecto\newpage         
%
\checkandfixthelayout
\setlength{\evensidemargin}{\oddsidemargin}
%
%
%
%
%
%
\frontmatter
\pagenumbering{roman}
\begin{titlingpage}
\begin{SingleSpace}
\calccentering{\unitlength} 
\begin{adjustwidth*}{\unitlength}{-\unitlength}
\vspace*{8mm}
\begin{center}
\rule[0.5ex]{\linewidth}{2pt}\vspace*{-\baselineskip}\vspace*{3.2pt}
\rule[0.5ex]{\linewidth}{1pt}\\[\baselineskip]
{\huge Implicit Media Tagging and Affect Prediction from video of spontaneous facial expressions, recorded with depth camera }
\\[4mm]
\rule[0.5ex]{\linewidth}{1pt}\vspace*{-\baselineskip}\vspace{3.2pt}
\rule[0.5ex]{\linewidth}{2pt}\\
\vspace{6.5mm}
{\large By}\\
\vspace{6.5mm}
{\Large\textsc{Daniel Hadar}}\\
\vspace{11mm}
{\large Under the supervision of}\\
\vspace{4.5mm}
{\large\textsc{Prof. Daphna Weinshall}}\\
\vspace{46mm}
\includegraphics[scale=0.1]{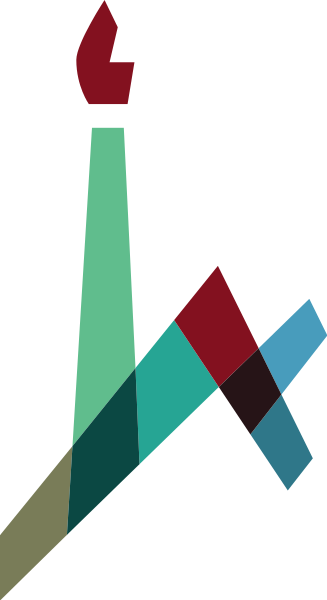}\\
\vspace{6mm}
{\large Department of Cognitive Science\\
\textsc{Hebrew University}}\\
\vspace{11mm}
\begin{minipage}{10cm}
A thesis submitted in partial fulfillment of the requirements for the degree of \textsc{Master of Cognitive Science} in the Faculty of Humanities.
\end{minipage}\\
\vspace{9mm}
{\large\textsc{December 2016}}
\vspace{12mm}
\end{center}
\begin{flushright}
\end{flushright}
\end{adjustwidth*}
\end{SingleSpace}
\end{titlingpage}
%
\chapter*{Abstract}
\begin{DoubleSpace}

We present a method that automatically evaluates emotional response from spontaneous facial activity recorded by a depth camera. The automatic evaluation of emotional response, or affect, is a fascinating challenge with many applications, including human-computer interaction, media tagging and human affect prediction. Our approach in addressing this problem is based on the inferred activity of facial muscles over time, as captured by a depth camera recording an individual's facial activity. Our contribution is two-fold: First, we constructed a database of publicly available short video clips, which elicit a strong emotional response in a consistent manner across different individuals. Each video was tagged by its characteristic emotional response along 4 scales: \emph{Valence, Arousal, Likability} and \emph{Rewatch} (the desire to watch again). The second contribution is a two-step prediction method, based on learning, which was trained and tested using this database of tagged video clips. Our method was able to successfully predict the aforementioned 4 dimensional representation of affect, as well as to identify the period of strongest emotional response in the viewing recordings, in a method that is blind to the video clip being watch, revealing a significantly high agreement between the recordings of independent viewers.

\end{DoubleSpace}
\clearpage

%
\chapter*{Acknowledgments}

\noindent
My humble thanks and appreciation to my supervisor, Prof. Daphna Weinshall, that guided me throughout this research and was deeply involved in it. I have been privileged to have her guidance and support.
\bigbreak
\noindent
In addition, I am grateful to Talia Granot, for her practical assistance regarding utilizing facial expressions, as well as our brainstorm meetings that provided me with new ideas and inspiration.
\bigbreak
\noindent
Many thanks also goes to the readers of this dissertation: Hillel Aviezer and Nir Fierstein, for their constructive comments.
\bigbreak
\noindent
Last but not least, I would like to thank my parents, Limor and Shuki, for their never-ending backing and support, and to my beloved wife Liat, for her endless optimism and encouragement.

\clearpage
%
%
\renewcommand{\contentsname}{Table of Contents}
\maxtocdepth{section}
\tableofcontents*
\addtocontents{toc}{\par\nobreak \mbox{}\hfill{\bf Page}\par\nobreak}
\clearpage
%
\listoftables
\addtocontents{lot}{\par\nobreak\textbf{{\scshape Table} \hfill Page}\par\nobreak}
\clearpage
%
\listoffigures
\addtocontents{lof}{\par\nobreak\textbf{{\scshape Figure} \hfill Page}\par\nobreak}
%
%
\mainmatter
\chapter{Introduction}
\label{chap:intro}

\initial{T}he Roman Philosopher Cicero wrote, \textsl{"The face is a picture of the mind"}. This statement has been discussed and debated repeatedly over the years, within the scientific community and outside it -- are the face really a window to the soul? Does human emotion reflect in facial expressions? There is a vast agreement among scholars that the answer is, at least partially -- yes. Hence, we asked the following question -- could it be done automatically? That is, could computer vision tools be utilized to evaluate humans' emotional state, given their facial expressions?


In the past two decades we had witnessed an increased interest in automatic methods that extract and analyze human's emotions, or \emph{affective state}. The potential applications of automatic affect recognition vary from human computer interaction to emotional media tagging, including for example the creation of a user's profile in various platforms, building emotion-driven HCI systems, and emotion-based tagging of dating sites, videos on YouTube or posts on Facebook. Indeed, in recent years media tagging has received much attention in the research community (\emph{e.g. }\cite{soleymani2012human, sariyanidi2015automatic}).

In this work we took advantage of the emerging technology of depth cameras. Recently, depth cameras based on structured light technology have emerged as a means to achieve effective human computer interaction in gaming, based on both gestures and facial expressions \cite{KinectSceletal}. We used a depth camera (Carmine 1.09) to record participants facial response to a set of video clips designed to elicit emotional response, and developed two types of pertinent prediction models for automatic quantitative evaluation of affect: models for tagging video clips from human facial expressions (\emph{i.e.} implicit media tagging), and models for predicting viewers affective state, given their facial behavior (\emph{i.e.} affect prediction). Respectively, a clear separation between them should be drawn.

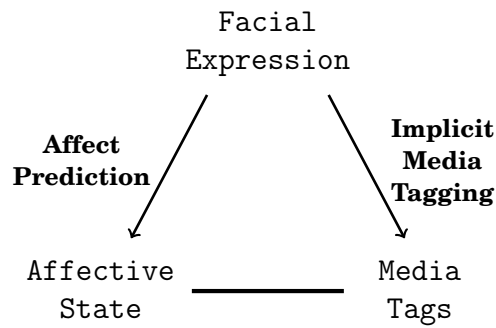
\begin{figure}[h]
  \centering
  \begin{tikzpicture}

  \node at (-0.2,0) {\ttfamily \large \begin{tabular}{c} Affective \\ State \end{tabular}};
  \node at (4,0) {\ttfamily \large \begin{tabular}{c} Media \\ Tags \end{tabular}};
  \node at (2,3.3) {\ttfamily \large \begin{tabular}{c} Facial \\ Expression \end{tabular}};

  \draw[line width=1.6pt] (1,0) -- (3,0);
  \draw[arrows=->,line width=1pt] (2.8,2.6) --(3.8,0.7);
  \draw[arrows=->,line width=1pt] (1.2,2.6) -- (0.2,0.7);
  
  \node at (4.3,1.7) {\small \bfseries \begin{tabular}{c}  Implicit \\  Media \\ Tagging \end{tabular}};
  \node at (-0.45,1.7) {\small \bfseries \begin{tabular}{c}  Affect \\  Prediction \end{tabular}};
  
  \end{tikzpicture}
  \caption{The triangular relationship between the facial expression, the media tags and the viewer's affective state.}
  \label{fig:intro_triangular_relationship}
\end{figure}

Both implicit media tagging and affect prediction concern the estimation of emotion-related indicators based on non-verbal cues, but they differ in their target: in the first, the purpose is predicting attributes of \emph{the multimedia stimuli}, while in the latter, \emph{the human affect} is the matter in hand. This distinction could be made clear by observing the triangular relationship between the video clip's affective tagging, the facial response to it and the viewer's reported emotional feedback (see Figure~\ref{fig:intro_triangular_relationship}) -- implicit media tagging concerns the automated annotation of a stimuli directly from spontaneous human response, while affect prediction deals with predicting the viewer's affective state. To be noted that objects and locations identification is not a part of this work's scope (\emph{e.g.} \cite{shotton2014image}), but only emotional-related tags.

As opposed to \emph{explicit} tagging, in which the user is actively involved in the tagging process, \emph{implicit} tagging is done passively, and relies only on the typical interaction the user have with such stimuli (\emph{e.g.} watching a video clip). As such, it is less time and energy consuming, and more likely to be free of biases. It has been suggested that explicit tagging tends to be rather inaccurate in practice; for example, users tend to tag videos according to their social needs, which yields tagging that could be reputation-driven, especially in a setup where the user's friends, colleagues or family may be exposed to their tags \cite{pantic2009implicit}.

\begin{figure}[htb]
  \centering
\includegraphics[width=0.45\textwidth]{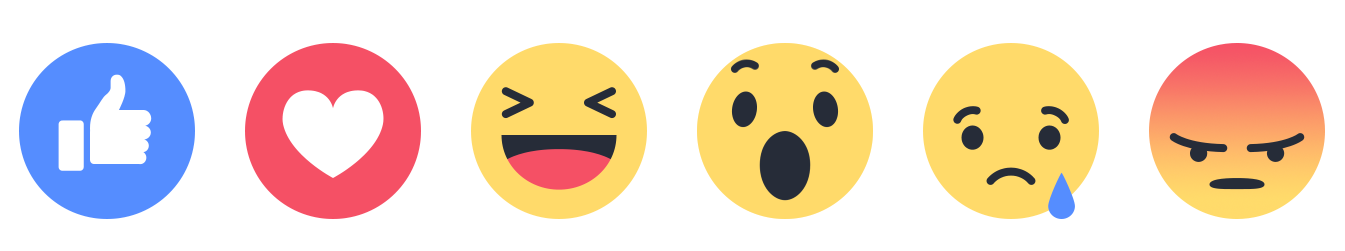} 
		\small	
		\caption{Facebook's reactions.\label{fig:fb_reactions}}
\end{figure}

In recent years, media tagging had became an integral part of surfing the internet. Many web platforms allow (and even encourage) users to label their content by using keywords (\emph{e.g.} \emph{funny}, \emph{wow}, \emph{LOL}) or designated scales (\emph{e.g.} Facebook's reactions, see Figure~\ref{fig:fb_reactions}). Clearly, non-invasive methods which produce such tags \emph{implicitly} can be of great interest. That being said, implicit media is complementary (rather than contradictory) to explicit media tagging, and as such can be used for assessing the correctness of explicit tags \cite{jiao2010implicit}, or for examining the inconsistency between the intentional (explicit) and involuntary (implicit) tagging \cite{mauss2009measures,maestas2016subject}.

Altogether, we learned four different models that share a similar inner-mechanism, but vary in their prior knowledge and target. Specifically, two models were trained to predict the \emph{clip's affective rating} (implicit media tagging), while two other models were trained to predict the \emph{viewer's subjective affective state} for each individual (affect prediction). Specifically, the first model predicts the \emph{affective rating} of a new clip given the facial expressions of a single viewer to a set of known clips, while the second model uses the facial expressions of several viewers. This can be useful for the affective tagging of new video clips by a system relying on the facial expressions of a group of viewers, who had previously been recorded viewing a set of clips with pre-determined affective rating. If no pre-determined affective ratings are available, the third model predicts the \emph{subjective rating} of a new clip, when given the facial expressions and subjective ratings of a single viewer to a set of clips which does not include the new clip. 

Affect computation from facial expressions cannot be readily separated from the underlying theories of emotional modeling, which rely on assumptions made by researchers in the field of psychology and sociology, and in some cases are still under debate. Thus the \emph{Facial Action Coding System} (FACS), originally developed by the Swedish anatomist Carl-Herman Hjortsj{\"o} \cite{hjortsjo1969man}, is often used to analyze facial activity in a quantitative manner. FACS gives a score to the activity of individual facial muscles called \emph{Action Units} (AUs), based on their intensity level and temporal segments. In our work we used the commercial software \emph{Faceshift} \cite{FaceShift} to extract quantitative measures of over 50 AUs from recordings of spontaneous facial activity.

Our goal was to achieve an automatic model that infers descriptive ratings of emotion. To this end we employed the \emph{dimensional approach} to modeling emotion, a framework whose elements are bipolar axes constituting the basis of a vector space (not necessarily orthonormal), and where it is assumed that every emotion can be described as a linear combination of these axes. We represented emotion by a combination of two key scales -- \emph{Valence} and \emph{Arousal} (see discussion in Section~\ref{sec:back_quantification}). In addition, we added 2 contemporary scales -- \emph{Likability} and \emph{Rewatch} (the desire to watch again), which are more suitable for modern uses in HCI and media tagging (following \cite{mcduff2014automatic}).

Our method is based on learning from a tagged database of video clips. In order to train a successful algorithm and be able to test it against some meaningful ground truth, we needed a database of video clips which can invoke strong emotional responses in a consistent manner across individuals. Since no reliable database of video clips that suites our needs exists at the moment, we constructed a new database from publicly available video clips (see discussion in Chapter~\ref{chap:db}). This database is available to the community and can be inspected in~\cite{eevdbgit}. A similar empirical procedure was used to collect data for the training and evaluation of our method, as described in Chapter~\ref{chap:method}.

Next, we developed a vector representation for videos of facial expressions, based on the estimated FACS measurements. Specifically, we started from AUs computed automatically by \emph{Faceshift} \cite{FaceShift} from a depth video. The procedure by which we have obtained a concise representation for each video of facial expressions, which involved quantification of both dynamic and spatial features of the AUs activity, is described in Section~\ref{sec:method_features}.

In Section~\ref{sec:method_models} we describe the method by which we learned to predict affect from the ensuing representation for each viewer and each viewed clip. In the first step of our method we generated predictions for small segments of the original facial activity video, employing linear regression to predict the 4 quantitative scales which describe affect (namely \emph{valence}, \emph{arousal}, \emph{likability} and \emph{rewatch}, \emph{a.k.a. VALR}). In the second step we generated a single prediction for each facial expressions video, based on the values predicted in the first step for the set of segments encompassed in the active part of the video. The results are described in Chapter~\ref{chap:results}, showing high correlation between the predicted scores and the actual ones.

There are three main novelties in this work are: first, our models are based directly on the automatically inferred activity of facial muscles over time, considering dozens of muscles, in a method which is blind to the actual video being watched by the subject (\emph{i.e.} the model isn't aware of any of the stimuli details and attributes); Second, the facial muscular activity is measured using only a single depth camera; And third, we present a new publicly available database of emotion eliciting short video clips, which elicit a strong affective response in a consistent manner across different individuals.
\clearpage

\let\textcircled=\pgftextcircled
\chapter{Theoretical Background and Previous Work}
\label{chap:back}

\initial{H}uman facial expressions are a fundamental aspect of our every-day lives, as well as a popular research field for many decades. It is not always thought-about or spoken-of, but many of the decisions we take are based upon other people's facial expressions, for example, deciding whether to ask someone's phone number ("is he smiling back?"), or whether a person is to be trusted. Psychologists have been claiming assiduously that human beings are in fact experts of facial expressions (as a car-enthusiast would be an expert of car models), in terms of being able to distinguish between highly similar expressions, recognize large amount of different faces (of friends, family, co-workers, etc.) and being sensitive to subtle facial gestures.

\section{Quantification of Emotion and Facial Expressions}
\label{sec:back_quantification}
For over a 100 years now, scholars have been interested in finding a rigorous model for the classification of emotion, seeking to express and explain every emotion by a minimal set of elements. The work in this field can be divided into two approaches: the \emph{categorical approach} and the \emph{dimensional approach} (see \cite{gunes2011emotion} for a comprehensive survey). These approaches are not contradictory, as they basically offer alternative ways to describe the same phenomenon. Briefly, the \emph{categorical approach} postulates the existence of several basic and universal emotions (typically surprise, anger, disgust, fear, interest, sadness and happiness), while the \emph{dimensional approach} assumes that each emotion can be described by a point in a $2D/3D$ coordinate system (typically valence, arousal and dominance).

In this work we collected emotional ratings over the dimensional approach, alongside free language descriptions, for several reasons: first, using \emph{forced choice paradigm} for emotion rating (compelling subjects to chose an emotion from a closed set of basic emotions) may yield spurious and tendentious results, that don't fully grasp the experienced emotion \cite{haidt1999culture}; second, keeping in mind that mixed feelings can arise from a single stimuli \cite{hemenover2007s,larsen2001can}, it follows that restricting participants to using a set of discrete emotions might bring about the loss of some emotional depth; and third, basic emotion tags could be extracted from valence and arousal ratings, as well as from the free language text \cite{aman2007identifying,danisman2008feeler}.

\begin{figure}[htb]
  \centering
\includegraphics[width=0.55\textwidth]{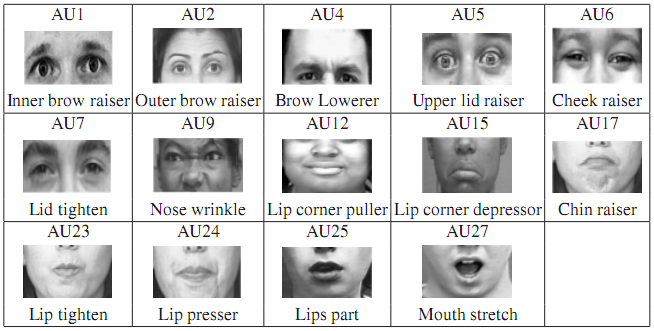} 
		\small	
		\caption{Examples from the Facial Action Coding System \cite{ekman1997face}\label{fig:AUlabeling_AU}.}
\end{figure}

In the 1970s Ekman and Friesen introduced FACS, that was developed to measure and describe facial movements \cite{ekman1978facial}. The system's basic components are Action Units (AUs), where most AUs are associated with some facial muscle, and some describe more general gestures, such as sniffing or swallowing (see examples in Fig.~\ref{fig:AUlabeling_AU}). Every facial expression can be described by the set of AUs that compose it. Over the years, Ekman and his colleagues expanded the set of AUs (and FACS expressiveness accordingly), adding head movements, eye movements and gross behavior. Currently there are several dozen AUs, over 50 of them are solely face-related. Describing a facial expression in terms of AUs is traditionally done by FACS-experts specifically trained for this purpose.

Automated FACS coding that will replace the manual one poses a major challenge to the field of computer vision \cite{pantic2007machine}. AUs extraction can be done using methods based on \emph{geometric features}, such as tracking points or shapes on the face, with features like position, speed and acceleration (e.g.,  \cite{pantic2004facial,pantic2006dynamics,gokturk2002model}), or using \emph{appearance based} methods based on changes in texture and motion of the skin, including wrinkles and furrows (e.g., \cite{littlewort2006dynamics,bartlett2006fully,bartlett2005recognizing,anderson2006real}). The use of geometric features tends to yield better results, since appearance based methods are more sensitive to illumination conditions and to individual differences, though a combination of both methods may be preferable \cite{tian2001recognizing}. A newer promising method is based on temporal information in AU activity, which was found to improve recognition as compared to static methods \cite{mavadati2014temporal, Koelstra2010}. Basic features are classified into AUs using model driven methods such as active appearance models (AAMs) \cite{lucey2007investigating} or data driven methods \cite{savran2012comparative}. While data driven methods require larger sets of data in order to cope with variations in pose, lightning and textures, they allow for a more accurate and person-independent analysis of AUs \cite{Savran2012}.

\section{Implicit Media Tagging and Affect Prediction}
\label{sec:back_imtandap}

To our knowledge, tagging of media implicitly and predicting viewers affective state based on data obtained from depth cameras is a recent and uncharted territory. A few exceptions include Niese \emph{et at.} \cite{niese2010emotion} who used evaluated 3D information from the raw data obtained by a 2D camera, and Tron \emph{et al.} \cite{tron2015automated, tron2016facial} who used depth cameras to classify the mental state of schizophrenia patients, based on their facial behavior.

\begin{figure}[!htb]
	\centering
	\includegraphics[width=0.80\textwidth]{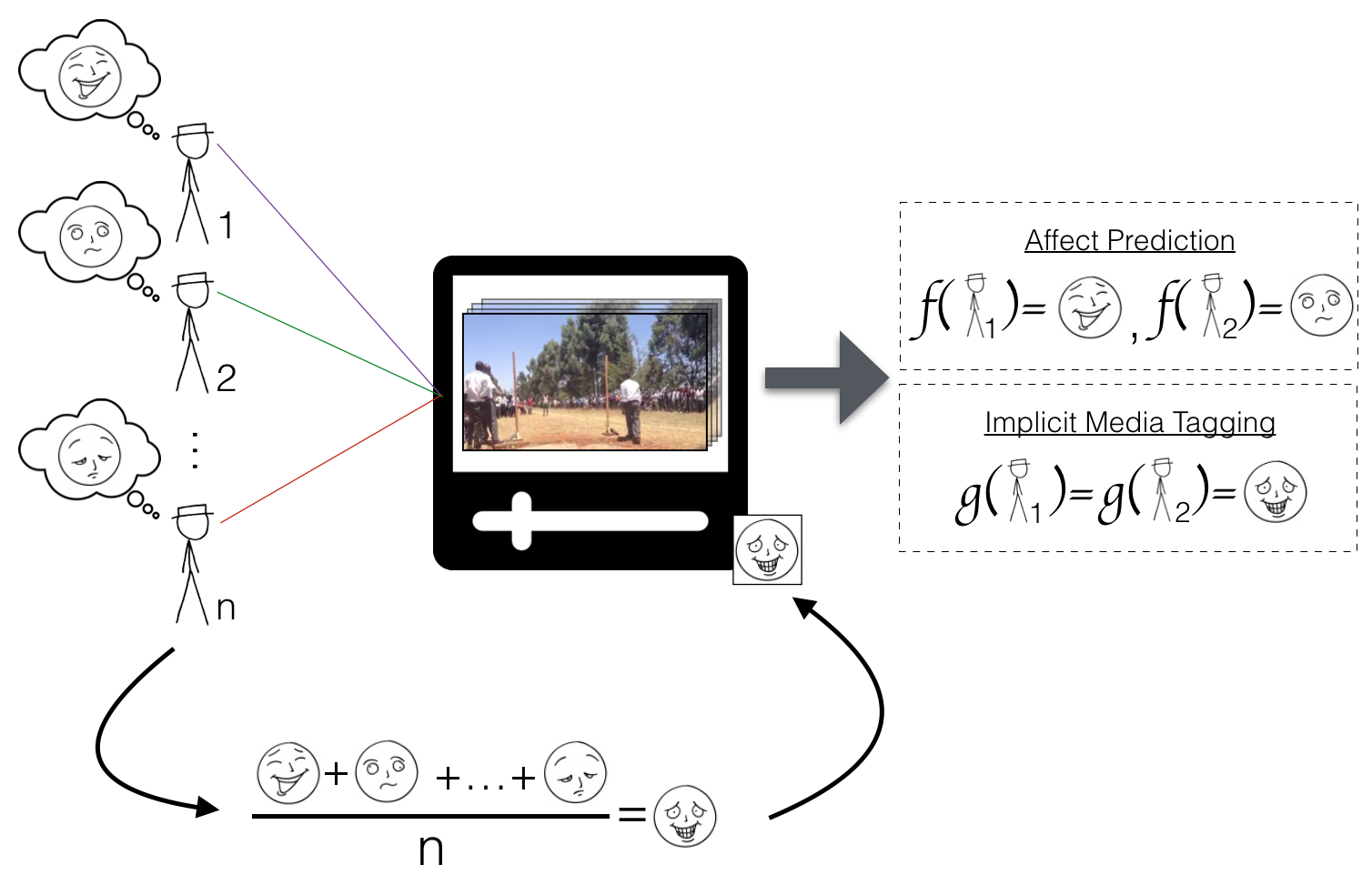} 
	\captionsetup{width=0.75\linewidth}
	\caption{Illustration of the difference between affect prediction (\(f\)) and implicit media tagging (\(g\)).
	\label{fig:functions}}
\end{figure}

The models in affect prediction are designed to predict each viewer's personal affective state, while in media tagging they attempt to predict a media tag based on input from (possibly different) viewers. Formally, for person \(p_i\) who is undergoing the affective state \(a_i\) while viewing the clip \(c_k\), an affect prediction model would be defined as: \(f(p_i)=a_i\), while an implicit media tagging model would be: \(g(p_i)=c_k\), and in particular, for viewers $i,j$ it holds that: \(f(p_i)=a_i\), \(f(p_j)=a_j\) (when $a_i$ isn't necessarily equals $a_j$), but \(g(p_i)=g(p_j)=c_k\). See illustration in Figure~\ref{fig:functions}.

\subsection{Implicit Media Tagging}
\label{subsec:back_imt}

A number of implicit media tagging theories and models have been developed based on different modalities, including facial expressions, low-level visual features of the face and body, EEG, eye gaze, fMRI, or audio-visual features of the stimuli.

\subsubsection*{Facial Expressions}

Hussein and Elsayed \cite{hussein2008studying} trained a classifier to predict the relevance of a document to a person based on her facial behavior, using 16 feature point along the face. Arapakis \emph{et al.} \cite{arapakis2009using} pursuit a similar goal, but the model's features included the level of basic emotions (e.g., how happy or angry the participant is), combined with other peripheral physiological metrics (such as skin temperature). They achieved higher success rates (accuracy of 66.5\% at best) and showed that a user's predicted emotional states can be used to predict the stimuli's relevance. 

Jiao and Pantic \cite{jiao2010implicit} suggested a facial expressions based model to predict the correctness of tags of images. They used 2D cameras to identify 19 facial markers (such as eyebrows boundaries, lip corners and nostrils) and used Hidden Markov Models to classify their movement along time. Their conclusion was that user's facial reactions convey information about the correctness of tags associated with multimedia data. Similarly, Tkalcic \emph{et al.} \cite{tkalcic2013affective} presented a method to predict valence, arousal and dominance tags for images, based on the user's facial expressions (given as a sequence of images).

As for video tagging, most works focus on affect prediction. Among the few that discuss media tagging are Zhao \emph{et al.} \cite{zhao2011video}, who proposed a method that based only on the participant's facial expressions to predict categories of movies (\emph{e.g.} comedy, drama and horror). Almost all papers that use facial expressions for implicit media tagging combine it with additional input, such as content from the video itself or additional physiological measures. Bao \emph{et al.} \cite{bao2013your} added the user's acoustic features, motion and interaction with the watching device (tablet or mobile). Wang \emph{et al.} \cite{wang2015implicit} used head motion as well as facial expressions, combined with "common emotions" (the emotions that are likely to be elicited from the majority of users) to predict video's tags, in terms of basic emotions. Another common measurement uses the EEG signal (\emph{e.g.} \cite{koelstra2013fusion, soleymani2016analysis}).

\subsubsection*{Other Methods}
The use of EEG in this field is relatively common, presumably because it is non-invasive and relatively cheap, permitting the appealing notion that one may be able to tap specific brain localization for different emotional states. EEG is popular for tagging of objects and landscapes (\emph{e.g.} \cite{gerson2006cortically,kapoor2008combining, koelstra2009eeg}), but it is also commonly used for emotion-related tagging.

Yazdani \emph{et al.} \cite{yazdani2009implicit} implemented an EEG-based system that predicts a media's tag, in terms of basic emotions \cite{ekman1992argument}; They achieved an average of 80.19\% over a self-developed database of 24 short video clips. Soleymani \emph{et al.} used EEG-based methods to tag clips from MAHNOB-HCI database \cite{soleymani2012multimodala}, once combined with pupillary response and eye gaze \cite{soleymani2012multimodalb} and once without \cite{soleymani2013multimedia}, and reached a success rate of \(F1=0.56\) for valence and \(F1=0.64\) for arousal.
Wang \emph{et al.} \cite{wang2014hybrid} expanded the battery of tools in use, adding features from the video stimuli itself to the EEG recordings, to create a fusion method that achieved an accuracy score of 76.1\% over valence and 73.1\% over arousal.

Han \emph{et al.} \cite{han2015arousal} proposed a method that predicts the arousal rating of video clips based on fMRI scans and low-level audio-visual features; this method achieved an average accuracy score of 93.2\% for 3 subjects. Jiang \emph{et al.} \cite{jiang2012music} proposed a framework that interweaves fMRI results to low-level acoustic features and thus enables audio tagging based on fMRI without actually using fMRI further on. Abadi \emph{et al.} \cite{abadi2013user} combined MEG signal and peripheral physiological signals (such as EOG and ECG) to predict the ratings of movies and music clips, including valence, arousal and dominance ratings, with accuracy score of 0.63, 0.62 and 0.59 (respectively) at best. Similar to \cite{bao2013your}, in \cite{silveira2013predicting} the authors also proposed a method to estimate movie's ratings, based instead on Galvanic Skin Response (GSR).

\subsection{Affect prediction}
\label{subsec:back_affectprediction}

Similarly to implicit media tagging, affect prediction models are usually composed of several inputs -- some physiological (\emph{e.g.} facial expressions, brain activity or acoustic signals) and some are derived from the media the participant is exposed to (\emph{e.g.} the video clip's tagging or visual and acoustic features). Moreover, they typically rely on the participant's reported emotional state. 

Lin \emph{et al.} \cite{lin2014fusion} combined EEG and acoustic characteristics from musical content to evaluate the participant's reported valence and arousal. Similarly, Chen \emph{et al.} \cite{chen2015electroencephalogram} used music pieces as a stimuli and combined EEG with subject's gender data to predict valence and arousal. Zhu \emph{et al.} \cite{zhu2014emotion} used video clips as stimuli, and also combined between EEG measurements and acoustic/visual features from the clips to predict valence and arousal. Lee \emph{et al.} \cite{lee2014emotion} had used a similar method, but utilized 3D fuzzy GIST for the features extraction and 3D fuzzy tensor for the EEG feature extraction, and predicted valence only. The last two had predicted only a binary result (positive or negative).

As for facial expressions, Soleymani \emph{et al.} \cite{soleymani2016analysis} compared it to EEG, and found that the results from facial expressions are superior to the results from EEG as a predictor for affective state, as most of the emotionally valuable content in EEG features is a result of facial muscle activity (but EEG signals still carry complementary information). McDuff \emph{et al.} \cite{mcduff2014automatic} used a narrow set of facial features (namely smiles and their dynamics) to predict participant's likability and desire to watch again of 3 Superbowl ads, with crowdsourced data collected from the participants webcams. Bargal \emph{et al.} \cite{bargal2016emotion} were among the first to use deep neural networks on facial expressions (captured by 2D camera) to predict affective states (in terms of basic emotions), and among the few to propose a model based only on facial expressions. In addition, several papers suggest methods to predict participant's level of interest based on their facial expressions, such as Peng \emph{et al.} \cite{peng2010real} that implemented a model based on head motion, saccade, eye blinks, and 9 raw facial components; they predicted an emotional score (as well as an attention score) and combined them to predict the level of interest. Arapakis \emph{et al.} \cite{arapakis2009enriching} proposed a multimodal recommendation system, and showed that the performance of user-profiling was enhanced by facial expression data.

\section{Depth Cameras}
The depth camera used in our study employs IR technology, in which infra-red patterns are projected onto the $3D$ scene, and depth is computed from the deformations created by $3D$ surfaces \cite{gu2009creating}. The technology enables to capture facial surface data, which is less sensitive to head pose and to lightning conditions than $2D$ data, and yields better recognition of AUs \cite{sandbach2012static,tron2016facial}. One drawback, however, is that the depth resolution is rather low, and therefore the image may contain small artifacts in highly reflective and non-reflective regions, or holes in regions not covered by the projector \cite{Savran2010}.

\section{Facial Response Highlight Period}
One substantial component of our models is identifying the most informative time frame in the participant's facial response. In fact, during most of the watching time, most participants' facial behavior showed little to no emotional response; therefore, we sought to find the part of the clip where relevant and informative activity was taking place. Money and Agius \cite{money2008video} distinguished between two types of highlight period localization techniques: internal and external; the first utilized information from the video itself, such as image analysis, objects identification and audio features, while the second analyses information which can be obtained irrespective of the video, such as viewer's physiological measures or contextual information, the time of the day, the device, or the location in which the viewer is watching the video.

Both internal and external techniques have been exploited to localize a video's highlight periods. Examples for internal localization are the work of Chan and Jones \cite{chan2005affect} that extracted affective labels (valence and arousal) from the video's audio content, and the work of Xu \emph{et al.} \cite{xu2014three} that proposed a framework based on the combination of low-level audiovisual features with more progressive features, such as dialogue analysis for informative keywords and emotional intensity.

Our highlight period localization technique is an external one, that is blind to the video clip, and is based solely on the viewers facial expressions (see Section~\ref{sec:method_hp} for details). Examples of external technique are the works of Joho \emph{et al.} \cite{joho2009exploiting,joho2011looking} and Chakraborty \emph{et al.} \cite{chakraborty2015using}. Joho \emph{et al.} localized highlight periods from viewers facial expressions, based on two feature types: expressions change rate, and pronunciation level -- the relative presence of expressions in each of three emotional categories: neutral, low (anger, disgust, fear, sadness) and high (happiness and surprise). Chakraborty \emph{et al.} harnessed highlight periods localization with a model composed of viewers facial expressions and heart rate, in order to detect sports highlights.
\clearpage

\let\textcircled=\pgftextcircled
\chapter{Database Construction}
\label{chap:db}

\initial{T}his chapter describes the database of emotion eliciting short video clips we have developed. There are currently several available databases of this kind, but unfortunately non of them is suited for our needs. In Section~\ref{sec:db_elicit} we review these databases and the relevant literature, and subsequently we list the criteria that our database must meet (Section~\ref{sec:db_criteria}). The full database can be found online \cite{eevdbgit}, also reviewed in Section~\ref{sec:db_overview}. The method we employed to collect ratings for the clips is very similar to the method we used for developing our model, and it is described in Section~\ref{sec:db_method} as well as in Chapter~\ref{chap:method}. The results and their analysis are described in Section~\ref{sec:db_res}, showing high inter-raters agreement.

\section{Eliciting Emotion via Video Clips}
\label{sec:db_elicit}
The traditional stimuli used to elicit emotion in human participants under passive conditions (without active participation by such means as speech or movement) are sound (e.g., \cite{zentner2008emotions}), still pictures (e.g., \cite{delaveau2016antidepressant, mcrae2012development}) and video clips (e.g., \cite{baveye2015liris}). Although still pictures are widely used, there are several limitation to this method. First, the stimulus is static, thus it lacks the possibility of examining the dynamics of an evoked emotion \cite{Zeng2009}. Second, IAPS \cite{lang2008international} is by far the most dominant database in use, and therefore almost every emotion-related experiment uses it. Video clips are also not free of issues, but it seems that they are the most suitable for eliciting strong spontaneous authentic emotional response \cite{frazier2004respiratory,palomba2000cardiac,westermann1996relative}. Thus, for our needs, an emotion eliciting database of video clips is required.

The first pioneers that developed and released affect tagged video clips databases were \cite{philippot1993inducing} and \cite{gross1995emotion}. Both used excerpts from well-known films such as \emph{Kramer vs. Kramer} and \emph{Psycho}, and collected ratings from human subjects about their experienced emotion. The considerable advances in emotional recognition in recent years had motivated scholars to develop and release large batteries of emotion eliciting video clips databases, and we review them in Appendix~\ref{app:app01}.

Initially we defined a set of criteria that our database must meet. However, each of the aforementioned databases lacks at least on of them (as could be seen in Appendix~\ref{app:app01}). We therefore followed the path taken by many other studies that preferred to collect and validate their own databases, or to modify profoundly an existing one (\emph{e.g. } \cite{tomarken1990resting,lisetti2004using,palomba2000cardiac,irie2010affective,teixeira2012determination,xu2008hierarchical,mcduff2015crowdsourcing}).

\section{Database Criteria}
\label{sec:db_criteria}
\begin{enumerate}[itemsep=1pt, topsep=12pt, partopsep=0pt]
\item \textbf{Duration. } Determining the temporal window's length depends on two main principles: (i) Avoid clips that elicit several distinct emotions in different times; (ii) Have the ability to use many different and diverse clips in a single experiment, without exhausting the subjects. Therefore the clips must be relatively short, but still long enough to elicit a strong clear emotional response. 
\item \textbf{VALR Rated. } A consequence of our choice of the \emph{dimensional approach} to describe emotion, alongside the scales of \emph{likability} and \emph{rewatch}.
\item \textbf{No Sensor Presence. } The awareness of subjects to being observed or recorded (\emph{i.e. } Hawthorne Effect) was found to possibly alter their behavior \cite{mccarney2007hawthorne,de2000impact,mccambridge2014systematic}.
\item \textbf{Diversity. } Clips should be taken from a variety of domains to reduce the effect of individual variability. Clearly the database should not contain only clips of cute cats in action or incredible soccer tricks, but a balanced mixture.
\item \textbf{Globally Germane. } Clips must be intelligible regardless of their soundtrack and content (\emph{e.g. } avoid regional jokes and tales).
\item \textbf{Unfamiliarity. } Clips should be such that uninformed viewers are not likely to be familiar with them on one hand, while being publicly available on the other hand.
\item \textbf{Not Crowdsourced. } Alongside its strengths, crowdsourcing can be problematic. For example, subjects are less attentive than subjects in a lab with an experimenter alongside them \cite{paolacci2010running}, and they differ in their psychological attributes (such as the level of self esteem) from other populations \cite{goodman2013data}. Moreover, Due to our use of depth cameras, the experiment must be held in a controlled environment (a depth camera is not a household item), and to keep correspondence between the experiment and the database ratings, it should also be tagged in the same environment.
\item \textbf{Publicly Available. } To encourage ratification of results and competition, the data must be accessible.
\end{enumerate}

\section{Overview}
\label{sec:db_overview}
The database is composed of 36 short publicly available video clips
(6-30 seconds, $\mu=20_{sec.}$). Each clip was rated by 26 participants on 5 scales, including valence, arousal, likability, rewatch (the desire to watch again) and familiarity, and in addition it was verbally described. Table~\ref{tab:db_overview} gives a summary of the database.

{\renewcommand{\arraystretch}{1.3}
\begin{table}[H]
	\begin{centering}
		\begin{tabular}{l>{\raggedright}p{5cm}}

\hline 
\textbf{Number of Clips} & {36}\tabularnewline
\textbf{Duration }{(in seconds)} & {6-30 ($\mu=20.0$)}\tabularnewline
\textbf{Number of Raters} & {26 (13 males and 13 females)}\tabularnewline
\textbf{Available Scales } & {Valence }{{[}1-5{]}}\tabularnewline
 & {Arousal }{{[}1-5{]}}\tabularnewline
 & {Likability }{{[}1-3{]}}\tabularnewline
 & {Rewatch }{{[}1-3{]}}\tabularnewline
 & {Familiarity }{{[}0, 1-2, 3+{]}}\tabularnewline
 & {Free Text }{{[}Hebrew{]}}\tabularnewline
\hline 

		\end{tabular}
	\par\end{centering}
	\caption{Emotion elicit database summary.}
	\label{tab:db_overview}
\end{table}}

\section{Method}
\label{sec:db_method}
A highly similar empirical framework was applied in both phases of data collection in this work, database construction and models development. Specifically, the depth camera was only present in the second phase. Here we only describe in details the clips selection process, and the methodology we adopted for the assessment phase is described in Chapter~\ref{chap:method}.

\paragraph{Clips Selection}
Over a 100 clips were initially selected from online video sources (such as YouTube, Vimeo and Flickr), to be eventually reduced to 36. We attempted to achieve a diverse set of unfamiliar clips, and therefore focused on lightly viewed ones. We excluded clips that might offend participants by way of pornography or brutal violence\footnote{As a rule of thumb, we used videos that comply with the YouTube's Community Guidelines \cite{youtubecommunityguidelines}.}. Several clips were manually curtailed to remove irrelevant content, scaled to fit a 1440 x 900 resolution, and balanced to achieve identical sound volume.

\paragraph{Clips Assessment}
26 volunteers with normal vision participated in the study, for which they received a small payment (13 males and 13 females between the ages 19-29, $\mu=23.5$). For the method employed, see Chapter~\ref{chap:method}.

\section{Results and Analysis}
\label{sec:db_res}
All clips were found to be significantly unfamiliar across all raters $(p<.0001)$, and no influence of gender was found. Moreover, ratings on all scales showed high inter-rater agreement with an average Intra-Correlation Coefficient (ICC) of $0.945$ (two-way mixed model, $CI=.95$). The results are illustrated in Figure~\ref{fig:db_results}, and detailed in Table~\ref{tab:db_results}.

{\renewcommand{\arraystretch}{1.3}
\begin{table}[H]
	\centering

	\begin{tabular}{|c|ccccc|cc|}
    	\hline
		 & Mean & Median & STD & \emph{min} & \emph{max} & ICC & $\alpha$ \tabularnewline
		\hline 
		\textbf{Valence} & 3.04 & 3.21 & 1.00 & 1.23 & 4.42 & 0.975 & 0.973 \tabularnewline
		\hline 
		\textbf{Arousal} & 3.08 & 3.02 & 0.65 & 1.73 & 4.19 & 0.926 & 0.923 \tabularnewline
		\hline 
		\textbf{Likability} & 2.02 & 2.04 & 0.56 & 1.12 & 2.92 & 0.954 & 0.952 \tabularnewline
		\hline 
		\textbf{Rewatch} & 1.73 & 1.75 & 0.44 & 1.08 & 2.54 & 0.927 & 0.924 \tabularnewline
		\hline 
    \end{tabular}
    
	\captionsetup{width=0.85\linewidth}
	\caption{Mean, median, standard deviation and range of the different scales over all clips, as well as Intra-Correlation Coefficient ($ICC$) and Cronbach's $\alpha$.}
	\label{tab:db_results}

\end{table}

There were strong correlations between several scales, most notably valence--likability (Pearson's $R=.92$), valence--rewatch ($R=.87$) and likability--rewatch ($R=.94$). Interestingly, no significant correlation was found between arousal and likability ($R=-.23$) or between arousal and rewatch ($R=-.04$); a possible explanation could be that some high arousal clips could be very pleasing (such as hilarious clips), while others are difficult to watch (like car accident commercials), as opposed to high valence clips that are unlikely to discontent anyone. As for valence--arousal,  a small negative correlation was found ($R=-.40$), possibly because clips with extremely high V-A values that mostly included pornographic content were excluded, although this result does correspond to prior findings \cite{gruhn2008age}. The results are shown in Figure~\ref{fig:db_corr}.

\begin{figure}[h]
	\centering
	\includegraphics[width=0.95\textwidth]{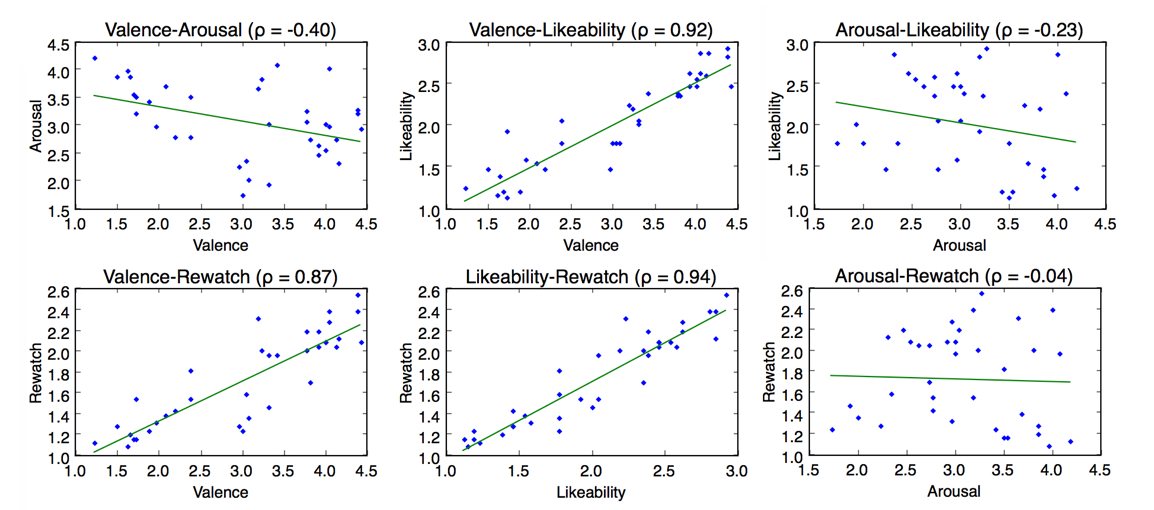} 
	\captionsetup{width=0.95\linewidth}
	\caption{Correlations between valence, arousal, likability and rewatch.
	\label{fig:db_corr}}
\end{figure}

\begin{figure}[t]
	\centering
	\includegraphics[width=1\textwidth]{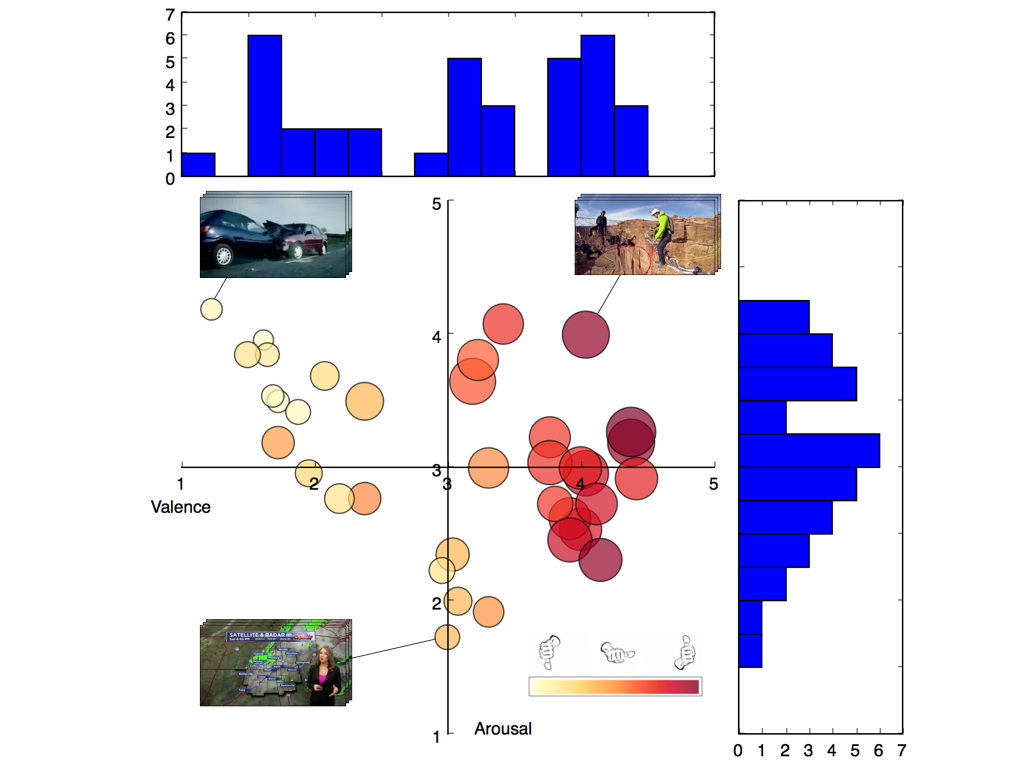} 
	\captionsetup{width=0.95\linewidth}
	\caption{Distribution of 4 subjective scores over all tested clips, where valence and arousal define the two main axes, also summarized in histogram form above and to the right of the plot. Size and color correspond respectively to the remaining 2 scores of rewatch and likability.
	\label{fig:db_results}}
\end{figure}
\clearpage

\let\textcircled=\pgftextcircled
\chapter{Method}
\label{chap:method}

\initial{D}ata collection had two phases: (i) Collect and evaluate a suitable database of video clips which elicit strong and consistent emotional responses in their viewers, as described in Chapter~\ref{chap:db}; (ii) Record peoples spontaneous facial expressions when viewing these clips. As mentioned, a highly similar empirical framework was used in both phases. In Section~\ref{sec:method_design} we describe the experimental environment and workflow. In Section~\ref{sec:method_recording} we describe the data collection process of the raw recordings, which was used later to calculate the features used in our models, and is elaborated in Section~\ref{sec:method_features}. Following the assemblage of features we learned our prediction models, a process that we describe in Section~\ref{sec:method_models}.

\section{Experimental Design}
\label{sec:method_design}
\paragraph{Participants} Data collection was carried out in a single room, well lit with natural light, with minimal settings (2 tables, 2 chairs, a whiteboard, 2 computers and no pictures hanging on the walls). Participants were university students recruited using banners and posters. 26 volunteers with normal vision (13 males and 13 females between the ages 19-29, $\mu=23.5$ in phase 1, 14 males and 12 females between the ages 20-28, $\mu=23.3$ in phase 2) participated in this study, for which they received a small payment.

\paragraph{Data Collection} Each data collection session consisted of the following stages (see Figure ~\ref{fig:method_exp_design}):
\begin{enumerate}
\item A fixation cross was presented for 5 seconds, and the participant was asked to stare at it.
\item A video clip was presented.
\item The participant described verbally his subjective emotion to the experimenter, using two sentences at most.
\item The participant rated her subjective feelings on a pre-printed ratings paper (following \cite{carvalho2012emotional}).
\end{enumerate}

Data collection was carried out via a self-written Matlab program, designed using Psychophysics Toolbox Extension \cite{brainard1997psychophysics,pelli1997videotoolbox,kleiner2007s}. Five discrete scales were used for rating: \emph{valence, arousal, likability, rewatch} (the desire to watch again) and \emph{familiarity}, alongside free text description in Hebrew. Specifically, we used SAM Manikins \cite{lang1995emotion} for valence and arousal, the "liking scale" for likability \cite{koelstra2012deap}, and self-generated scales for rewatch and familiarity (see Figure~\ref{fig:method_exp_design}). The SAM Manikins method was chosen because it is known to be parsimonious, inexpensive and quick, as well as comprehensive \cite{bradley1994measuring}. After every 4 trials, a visual search task was presented (\emph{``Where's Waldo?''}) in order to keep the participants focused, and to force them to change their head situs, sitting position and focal length. The clips order was randomized, and the entire procedure lasted for about an hour. We  encouraged participants to rate the clips according to their perceived, inner and subjective emotion. In addition, each participant completed the \emph{Big-Five Personality Traits} questionnaire \cite{john1999big}.

\begin{figure}[t]
	\centering
	\includegraphics[width=1\textwidth]{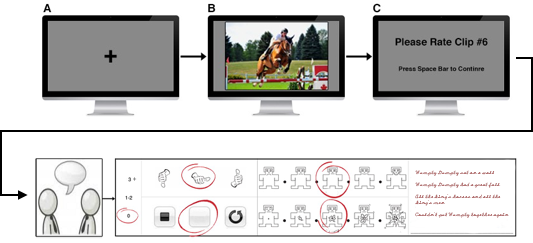} 
	\captionsetup{width=0.95\linewidth}
	\caption{The Experimental design.
	\label{fig:method_exp_design}}
\end{figure}

\section{Facial Expressions Recording}
\label{sec:method_recording}
18 of the 36 clips were selected from the database. Aiming for a diverse corpus, we chose clips whose elicited response spanned the spectrum of \emph{VALR} as uniformly as possible, also favoring clips with high intra-rater agreement.

Each participant's facial activity was recorded during the entire procedure, using a $3D$ structured light camera (Carmine 1.09). Participants were informed of being recorded and signed a consent form. 21 of the 26 participants reported after the experiment their belief that they were not affected by the recording, and that in fact they had forgotten of being recorded. Moreover, the subjective reports in the second phase on all 4 scales had a very similar distribution to the reports in the first phase: Valence ($R=.98,\ p<.0001$), Arousal ($R=.90,\ p<.0001$), Likability ($R=.97,\ p<.0001$) and Rewatch ($R=.95,\ p<.0001$). We therefore believe that the video affect tagging obtained in the database assemblage phase remains a reliable predictor of the emotional response elicited in the recorded experiment as well.

\section{Features}
\label{sec:method_features}
Previous work in this field calculated facial features either by extracting raw movement of the face without relating to specific facial muscles (e.g., \cite{wang2015implicit}), or by extracting the activity level of a single or a few muscles (e.g., \cite{mcduff2014automatic, yang2014zapping, vandal2015event}). In this work we extracted the \emph{intensity signals} of over 60 AUs and facial gestures; this set was further analyzed manually to evaluate tracking accuracy and noise levels. Eventually 51 of this set of AUs were selected to represent each frame in the clip for further analysis and learning, including eyes, brows, lips, jaw and chin movements (see example in Fig.~\ref{fig:method_clip_subj_au}).

Using this \emph{intensity level} representation, providing a time series of vectors in ${\mathbb{R}}^{51}$, we computed higher order features representing the facial expression more concisely. This set of features can be divided into 4 types: \emph{Moments}, \emph{Discrete States}, \emph{Dynamic} and \emph{Miscellaneous}.

\begin{description}[itemsep=0pt, topsep=12pt, partopsep=0pt]
\item{\textbf{Moments.} } The first 4 moments (mean, variance, skewness and kurtosis) were calculated for each AU in each facial video recording.

\item{\textbf{Discrete States Features.} } For each AU separately, the raw intensity signal was quantized over time using \emph{K-Means} ($K=4$), and the following four facial activity characteristic features were computed (see Figure~\ref{fig:method_discrete} for an example):

\begin{itemize}[itemsep=1pt, topsep=4pt, partopsep=0pt]
\item \emph{\textbf{Activation Ratio:}} Proportion of frames with any AU activation.
\item \emph{\textbf{Activation Length:}} Mean number of frames for which there was continuous AU activation.
\item \emph{\textbf{Activation Level:}} Mean intensity of AU activation.
\item \emph{\textbf{Activation Average Volume:}} Mean activation level of all AUs, was computed once for each expression. 
\end{itemize}

\begin{figure}[h]
 \centering
 
  \subbottom[][]{\includegraphics[height=.2\textheight]{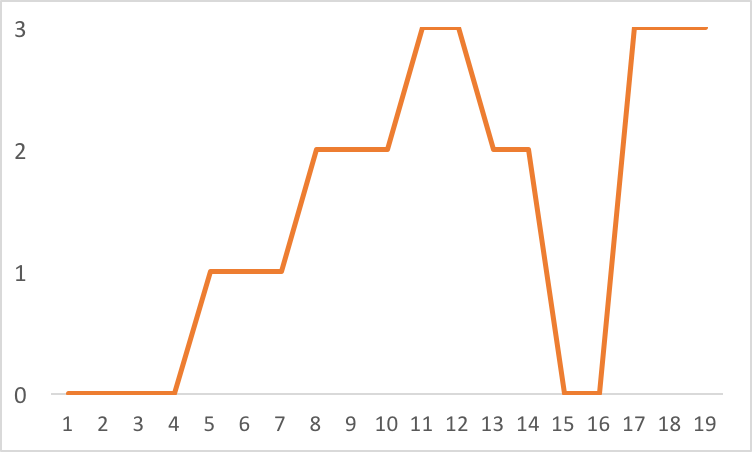}}\quad
  \subbottom[][]{\includegraphics[height=.2\textheight]{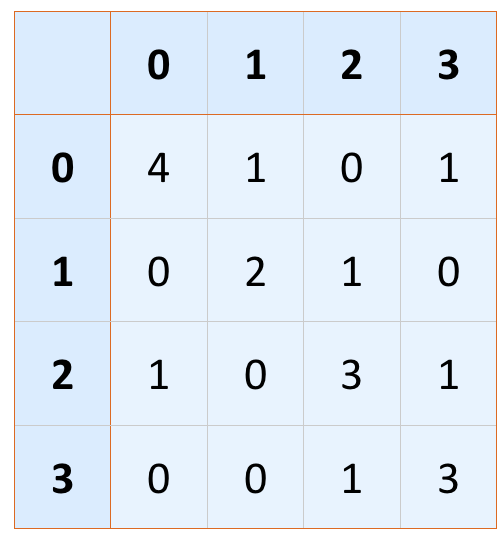}}\\

\caption{(A) Quantized AU signal ($K=4$), and (B) Its corresponding transition matrix. The number of frames labeled \emph{0,1,2,3} is 6,3,5,5, respectively. Therefore: $Activation Ratio=\frac{13}{19}$, $Activation Length=\frac{7}{19}$ and $Activation Level=1.473$, and  $ChangeRatio=\frac{6}{18}$, $SlowChangeRatio=\frac{4}{18}$, $FastChangeRatio=\frac{2}{18}$. }
\label{fig:method_discrete}
\end{figure}

\item{\textbf{Dynamic Features.} } A transition matrix $M$ was generated, measuring the number of transitions between the different levels described above, and three features were calculated for each AU based on it (see Figure~\ref{fig:method_discrete} for an example):

\begin{itemize}[itemsep=1pt, topsep=4pt, partopsep=0pt]
\item \emph{\textbf{Change Ratio:}} Proportion of transitions with level change.
\item \emph{\textbf{Slow Change Ratio:}} Proportion of small changes (difference of 1 quantum).
\item \emph{\textbf{Fast Change Ratio:}} Proportion of large changes (difference of 2 quanta or more).
\end{itemize}

\item{\textbf{Miscellaneous Features.} } Including the number of smiles and blinks in each facial response. The amount of smiles was calculated by taking the maximum of the amount of peaks in the signals of both lip corners, where \emph{peak} is defined as a local minimum which is higher by at least 0.75 as compared to its surrounding points. The amount of blinks was calculated in a similar manner, with a threshold of 0.2.

\end{description}

\begin{figure}[t]
	\centering
	\includegraphics[width=1\textwidth]{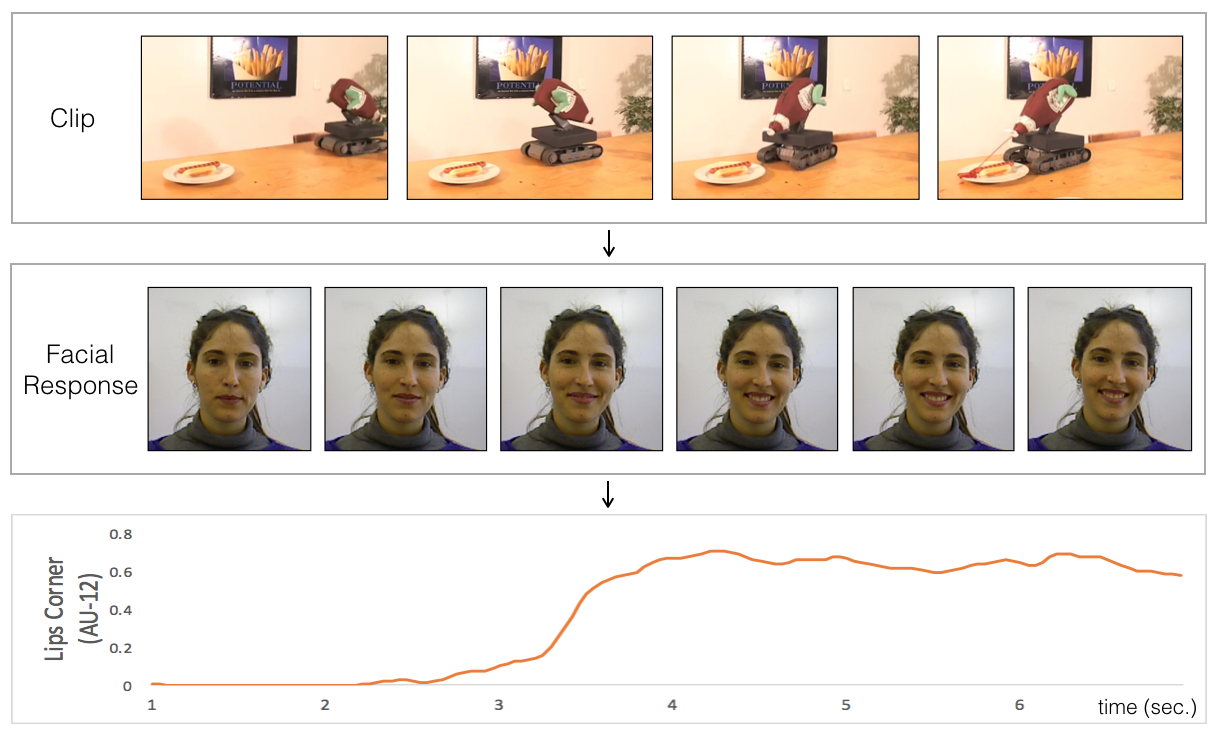} 
	\captionsetup{width=0.95\linewidth}
	\caption{Facial response (middle row) to a video clip (illustrated in the top row), and the time varying intensity of AU12 (bottom row).
	\label{fig:method_clip_subj_au}}
\end{figure}

\subsection*{Highlight Period}
\label{sec:method_hp}
In most video clips, during most of the viewing time, participants' facial activity showed almost no emotional response. Respectively, considering the entire duration of the facial expression when calculating features is not only unnecessary, but could actually harm the model as it adds noise to it. We therefore sought to find the time frame of each facial expression in which the relevant activity (in terms of affective response) was taking place. Specifically, we implemented a model that localized the \emph{highlight period} solely from the viewer's facial expression, in a technique that is blind to the video clip.

For each participant and clip, our model receives his/hers muscular intensity levels for the clip's duration (with 6 seconds margins from its beginning and end), and isolates the activity of gestures we found to be most informative (namely smiles, blinks, mouth dimples, lips stretches and mouth frowns). Later it localizes the 6 seconds window in which these gestures achieves maximal average intensity and variance. Excluding moments, All features were computed based only on the highlight period. Notice that for some clip \(C_i\), the highlight period might be different for every pair of subjects (although its duration will be the same, as it is an simplifying assumption of our model).

\section{Predictive Models}
\label{sec:method_models}

In this final step we learned two types of prediction models -- implicit media tagging models (\emph{IMT}), and affect prediction models (\emph{AP}). Both model types predict an affective rank (in VALR terms) when given as input a facial expression recording, represented by a vector in ${\mathbb{R}}^d$ feature space ($d=462$). Since the number of participants in our study was only 26, a clear case of small sample, the full vector representation -- if used for model learning -- would inevitably lead to overfit and poor prediction power. We therefore started by significantly reducing the dimension of the initial representation of each facial activity using PCA. Our final method employed a two-step prediction algorithm (see illustration in Figure~\ref{fig:2_step_methid}), as follows:

\begin{description}[itemsep=0pt, topsep=8pt, partopsep=0pt]
\item{\textbf{First step} } After the highlight period of each clip was detected (for each subject), it was divided into $n$ fixed size overlapping segments. A feature vector was calculated for each segment, and a linear regression model (\(f_1\)) was trained to predict the 4-dimensional affective scores vector of each segment.

\item{\textbf{Second step} } Two indicators (mean and std) of the set of predictions over all segments in the clip were calculated, and another linear regression model (\(f_2\)) was trained to predict the 4-dimensional affective scores from these indicators.
\end{description}

\begin{figure}[htb]
\centering
\includegraphics[width=0.8\textwidth]{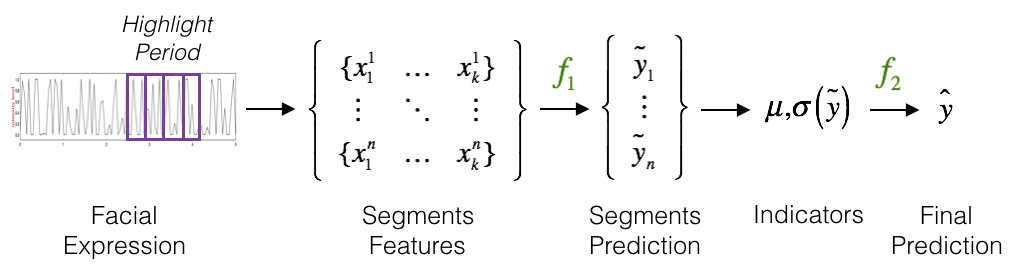}
\caption{Illustration of the two-step prediction algorithm.}
\label{fig:2_step_methid}
\end{figure}

Several parameters control the final representation of each facial expression clip, including the number of segments, the length of each segment, the percent of overlap between the segments, the final PCA dimension and whether PCA was done over each feature type or over all features combined. The values of these parameters were calibrated using cross-validation: given a training set of $l$ points, the training and prediction process was repeated $l$ times, each time using $l-1$ points to train the model and predict the value of the left out point. The set of parameters which achieved the best average results over these $l$ cross validation repetitions was used to construct the final facial expression representation of all datapoints.

\subsection*{Different types of predictive models}
Altogether, we learned four different models that shared this mechanism, but varied in their prior knowledge and target (see Table~\ref{tab:method_models_sum} for formal definition). Specifically, the first two models were trained to predict the \emph{clip's affective rating} as stored in the database (implicit media tagging), while the last two models were trained to predict the \emph{viewer's subjective affective state} for each individual (affect prediction).

\begin{description}[itemsep=0pt, topsep=12pt, partopsep=0pt]
\item{\textbf{Implicit media tagging of unseen clips (IMT-1).} }
This model is built using the facial expressions of a single viewer. Given the facial response of this viewer to a \textbf{new clip}, the model predicts the \emph{clip's affective rating}.

\item{\textbf{Implicit media tagging of unseen clips via multiple viewers (IMT-2).} }
Given the facial response of \emph{a set of familiar viewers} to a \textbf{new clip}, the model predicts the \emph{clip's affective rating}. Generalizing the first model, this second model predicts the new clip's affective rating by taking into account the prediction of \emph{all} viewers.


\item{\textbf{Viewer's affect prediction for an unseen clip (AP-1).} }
This model is built using the facial expressions of a single viewer. Given the facial response of this viewer to a \textbf{new clip}, the model predicts the \emph{viewer's subjective affective state}.

\item{\textbf{Affect prediction of new viewers (AP*).} }
This model is built separately for each clip, using the facial expressions of all the viewers who had watched this clip. Given the facial response of a \textbf{new viewer}, the model predicts the new viewer's \emph{subjective affective state} when viewing this clip. 

\end{description}

\begingroup
\setlength{\parindent}{0pt}
Formally, denote the following: 
\endgroup

\begin{itemize}[itemsep=0pt, topsep=8pt, partopsep=0pt]

\item{Let \(e_i^j\) denote the facial expression representation of viewer \(i\) to clip \(j\).}

\item{Let \(v_i^j\) (\emph{respectively:} \(a_i^j\), \(l_i^j\), \(r_i^j\)) denote the valence (\emph{respectively:} arousal, likability, rewatch) score of viewer \(i\) to clip \(j\).}

\item{Let \(s_i^j\equiv\left(v_i^j, a_i^j, l_i^j, r_i^j\right)\) denote the affective vector score of viewer \(i\) to clip \(j\).}

\item{Let \(\bv^j\) (\emph{respectively:} \(\ba^j\), \(\bl^j\), \(\br^j\)) denote the valence (\emph{respectively:} arousal, likability, rewatch) rating of clip \(j\).}

\item{Let \(\bc^j\equiv\left(\bv^j, \ba^j, \bl^j, \br^j\right)\) denote the affective vector rating of clip \(j\).}

\end{itemize}

{\renewcommand{\arraystretch}{1.6}
\begin{table}[H]
	\centering

	\begin{tabular}{|c||c|c|c|m{5cm}|}
    
    	\hline
		 \textbf{Model Name} & \textbf{Input} & \textbf{Model's supervision} & \textbf{Output} & \multicolumn{1}{c|}{\textbf{Comment}} \tabularnewline
		\hline 
        
		IMT-1 
        & $e_{m}^{k}$
        & $\begin{array}{c} \left\{e_m^j\right\}_{j=1,j \neq k}^{18} \\ \bc^{1}...\widehat{\bc^{k}}...\bc^{18} \end{array}$
        & $\bc^{k}$
        & A unique model is built for each viewer $m$. Note that the model is utterly unfamiliar with clip $k$.
        \tabularnewline \hline 
        
   		IMT-2 
        & $\left\{e_{i}^{k}\right\}_{i\in S,~S\subset[26]}$
        & $\begin{array}{c} \left\{\left\{e_i^j\right\}_{j=1,j\neq k}^{18}\right\}_{i=1}^{26} \\ \bc^{1}...\widehat{\bc^{k}}...\bc^{18} \end{array}$  
        & $\bc^{k}$
        & The output is a single model, which predicts the clip's rank ($\bc^k$) based on the average of the $m$ models computed in IMT-1. 
        \tabularnewline \hline
        
        
		AP-1 
        & $e_{m}^{k}$
        & $\left\{\left(e_m^j,s_m^j\right)\right\}_{j=1,j \neq k}^{18}$ 
        & $s_{m}^{k}$ 
        & Similarly to IMT-1, a unique model is built for each viewer $m$. The model is also unfamiliar with clip $k$.
        \tabularnewline \hline 
        
		AP* 
        & $\left\{e_{m}^{j}\right\}_{j=1}^{18}$
        & $\left\{\left\{\left(e_i^j,s_i^j\right)\right\}_{j=1}^{18}\right\}_{i=1,i\neq m}^{26}$ 
        & $\left\{s_{m}^{j}\right\}_{j=1}^{18}$
        & The model predicts $s_m^j$ separately for each clip $j$, and is also utterly unfamiliar with viewer $m$.
        \tabularnewline \hline 
        
    \end{tabular}
    
	\captionsetup{width=0.85\linewidth}
	\caption{Summary of models learned.}
	\label{tab:method_models_sum}

\end{table}

\clearpage

\let\textcircled=\pgftextcircled
\chapter{Results and Analysis}
\label{chap:results}

\initial{T}o evaluate the predictive power of our models, we divided the set of recordings following a Leave-One-Out (LOO) procedure. Specifically, for IMT-1, IMT-2 and AP-1 we trained each model based on \(n-1\) clips (\(n=18\)) and tested our prediction on the clip which was left out. For AP* the procedure was identical, up to  leaving a viewer out (instead of a clip), hence \(n=26\). The results are shown in Section~\ref{sec:results_performance}, followed by an analysis of the relative importance of the different feature types (moments, discrete state features, dynamic features and miscellaneous features) in Section~\ref{sec:results_feat_importance}, and the localization of highlight period in Section~\ref{sec:results_hp}.

\section{Learning Performance}
\label{sec:results_performance}
Learning performance was evaluated by \emph{Pearson's R} between the actual VALR scores and the models' predicted ones (see example in Figure~\ref{fig:corr_example}). Table~\ref{tab:results} shows the average VALR results over all clips/viewers (all correlations are significant, \emph{$p<0.0001$}).

\begin{figure}[htb]
\centering
\includegraphics[width=0.5\textwidth]{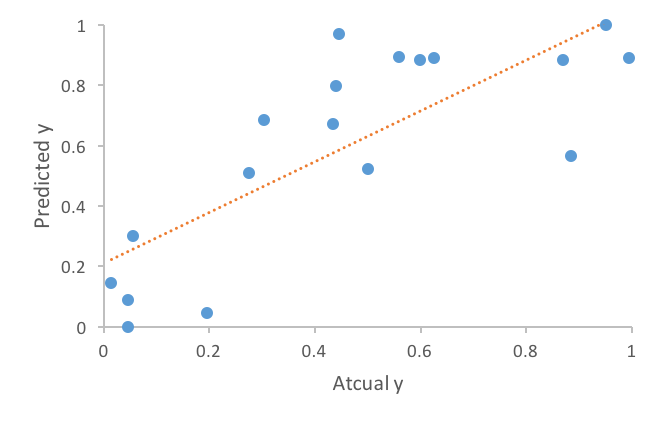} 
		\small	
		\caption{Correlation example between predicted and actual ratings of a single viewer's valence score (\emph{R=0.791}).
        \label{fig:corr_example}}
\end{figure}

{\renewcommand{\arraystretch}{1.3}
\begin{table}[ht]
	\centering

	\begin{tabular}{ccccc}
		 & \textbf{Valence} & \textbf{Arousal} & \textbf{Likability} & \textbf{Rewatch}  \tabularnewline
         \hline \hline
		
        \textbf{IMT-1} & .752 \scriptsize{(.14)} & .728 \scriptsize{(.07)} & .637 \scriptsize{(.22)} & .661 \scriptsize{(.15)}  \tabularnewline
		\hline 
        
        \textbf{IMT-2} & .948 \scriptsize{(.22)} & .874 \scriptsize{(.22)} & .951 \scriptsize{(.17)} & .953 \scriptsize{(.19)}  \tabularnewline
		\hline
		        
        \textbf{AP-1} & .661 \scriptsize{(.17)} & .638 \scriptsize{(.19)} & .380 \scriptsize{(.16)} & .574 \scriptsize{(.19)}  \tabularnewline
		\hline 
        
        \textbf{AP*} & .561 \scriptsize{(.26)} & .138 \scriptsize{(.21)} & .275 \scriptsize{(.23)} & .410 \scriptsize{(.17)}  \tabularnewline
		\hline
        
        \textbf{Report/Tags} & .783 \scriptsize{(.08)} & .461 \scriptsize{(.33)} & .659 \scriptsize{(.13)} & .561 \scriptsize{(.23)}  \tabularnewline
		\hline 
    \end{tabular}
    
	\captionsetup{width=0.8\linewidth}
	\caption{Mean (and std) of Pearson's R between the predicted and actual ranks, as well as the average correlation between the subjective report and media tags.}
	\label{tab:results}

\end{table}

Notably, implicit media tagging models reached higher success rates than the affect prediction ones. In particular, although IMT-1 and AP-1 both predict an affective state when given a familiar viewer's facial response to a new clip, the first yields noticeably higher success rates ($\mu=.131$). Based on these results, it could be suggested is that it is easier to predict the \emph{expected} affective state from viewer's facial expressions than the \emph{reported} one. This is rather surprising, as the opposite sounds more likely -- that one's facial behavior would be a better predictor to his/hers \emph{subjective} emotion than their \emph{expected} one. Under the assumption that the participants in this research proclaimed the true experienced emotion they underwent (as requested), it implies that human facial expressions are more faithful to the actual inner emotion than to the one reported. One reservation is that the viewer's subjective report is given on a discrete scale, while the media is tagged on a continuous one (as these are average ratings), thus the correlation is likely to be higher for the latter. This problem is averted by comparing the results after binarizing the predictions. 

\subsection*{Binary prediction}
\label{sec:results_binary}
Apart from the aforementioned reason, often what is needed in real-life applications is a discrete binary prediction rather than a continuous grade, in order to indicate, for example, whether the viewer likes a video clip -- or not. To generate such a measure, we binarized the actual ratings and the predicted ones, using the corresponding mean of each measure as a threshold, and calculated the accuracy score (ACC) between them. Since the scores around the mean are rather ambiguous, we eliminated from further analysis the clips whose original tag was uncertain. This included clips with scores in the range $\mu\pm \sigma$, where $\mu$ denotes the average score over all clips and $\sigma$ its $std$, thus eliminating $15\%$ on average of all data points. The results can be found in Table~\ref{tab:bin_results}.

{\renewcommand{\arraystretch}{1.3}
\begin{table}[ht]
	\centering

	\begin{tabular}{ccccc}
		 & \textbf{Valence} & \textbf{Arousal} & \textbf{Likability} & \textbf{Rewatch}  \tabularnewline
         \hline \hline
		
        \textbf{IMT-1} & 71\% \scriptsize{(.15)} & 56\% \scriptsize{(.14)} & 68\% \scriptsize{(.12)} & 73\% \scriptsize{(.13)}  \tabularnewline
		\hline
        
        \textbf{IMT-2} & 94\% \scriptsize{(.20)} & 90\% \scriptsize{(.18)} & 91\% \scriptsize{(.21)} & 91\% \scriptsize{(.22)}  \tabularnewline
		\hline
		
        
        \textbf{AP-1} & 70\% \scriptsize{(.17)} & 53\% \scriptsize{(.22)} & 49\% \scriptsize{(.18)} & 49\% \scriptsize{(.27)}  \tabularnewline
		\hline 
        
        \textbf{AP*} & 64\% \scriptsize{(.22)} & 50\% \scriptsize{(.17)} & 40\% \scriptsize{(.25)} & 42\% \scriptsize{(.26)}  \tabularnewline
		\hline 
    \end{tabular}
    
	\captionsetup{width=0.8\linewidth}
	\caption{Accuracy (and std) of the derived binary measure.}
	\label{tab:bin_results}

\end{table}

Comparing the results of IMT-2 to published state-of-the-art methods (\emph{e.g.} \cite{mcduff2014automatic}) demonstrates great success, as an ability to tag media in VALR terms with accuracy rates ranging around $90\%$ is unprecedented. That being said, it is important to note that these results were obtained using a newly composed database (presented in Chapter~\ref{chap:db}), therefore further research must be carried out for balanced comparison between methods.

Regardless to whether the scale is discrete, the ratio between the analogous algorithms (namely IMT-1 and AP-1) remains similar. Hence, these results support the aforementioned claim that human facial behavior is more faithful to the actual inner emotion than to the one reported. Furthermore, facial expressions allow for better predictions of media tags than the viewer's subjective rating, as the success rates of IMT-1 are generally higher than the correlation between the viewer's reported affective state and the media tags (see \emph{Report/Tags} is Table~\ref{tab:results}). These findings are supported by theories of emotional self-report bias, that could arise from many factors (such as cultural stereotypes, the presence of an experimenter, and the notion of the reports being made public).

Unfortunately, the binary predictions of AP* are generally no better than random predictions. Furthermore, for continuous ranks, the algorithm's predictive power is limited; when a model supervises a group of viewers (like AP*), it is preferable to predict a new viewer's affective state using only their \emph{affective reports}, without even relying on their facial behavior, as the \emph{average} affective rank of $n-1$ viewers provides a more accurate prediction to the $n$-th viewers (namely One-Viewer-Out method), as could be seen in Table~\ref{tab:results_ap_star_analysis}. In other words, when given a group of viewers' facial expressions and affective ranks, using the average of their ranks would yield better prediction for a new viewer's rank than his/hers facial expression.

{\renewcommand{\arraystretch}{1.3}
\begin{table}[h]
	\centering

	\begin{tabular}{ccccc}
		 & \textbf{Valence} & \textbf{Arousal} & \textbf{Likability} & \textbf{Rewatch}  \tabularnewline
         \hline \hline
        
        \textbf{AP*} & .561 \scriptsize{(.26)} & .138 \scriptsize{(.21)} & .275 \scriptsize{(.23)} & .410 \scriptsize{(.17)}  \tabularnewline
		\hline
        
        \textbf{Average ($n-1$)} & .779 \scriptsize{(.08)} & .472 \scriptsize{(.29)} & .645 \scriptsize{(.13)} & .552 \scriptsize{(.21)}  \tabularnewline
		\hline 
    \end{tabular}
    
	\captionsetup{width=0.8\linewidth}
	\caption{Mean (and std) of Pearson's R between the predicted and actual ranks of AP* and One-Viewer-Out method.}
	\label{tab:results_ap_star_analysis}

\end{table}

In addition, although seemingly AP-1 yields rather accurate predictions, deeper inspection reveals that an alternative mechanisms based on IMT-1 could be suggested, namely return the predictions of IMT-1 instead; this mechanism yields more accurate predictions \emph{of the subjective affect rank} than AP-1. In other words, if the media's tags are also available, it's preferable to train a model to predict these tags and rely on this model alone, rather than on a model trained to predict the viewer's subjective rank. Moreover, if other viewers data is also available, relying on it (namely returning IMT-2's predictions) is even more superior. We demonstrate this observation by inspecting the mean error (ME) between the actual subjective ranks and the predictions given by  the AP-1, IMT-1 and IMT-2 models, see Table~\ref{tab:results_ME}.

{\renewcommand{\arraystretch}{1.3}
\begin{table}[h]
	\centering

	\begin{tabular}{ccccc}
		 & \textbf{Valence} & \textbf{Arousal} & \textbf{Likability} & \textbf{Rewatch}  \tabularnewline
         \hline \hline
        
        \textbf{AP-1} & .265 & .370 & .388 & .371 \tabularnewline
		\hline
        
        \textbf{IMT-1} & .214 & .262 & .325 & .341 \tabularnewline
		\hline 
        
        \textbf{IMT-2} & .179 & .239 & .295 & .306 \tabularnewline
		\hline 
        
    \end{tabular}
    
	\captionsetup{width=0.8\linewidth}
	\caption{Mean error of all viewers subjective ranks and AP-1, IMT-1 and IMT-2's predictions.}
	\label{tab:results_ME}

\end{table}

\section{Relative Importance of Features} 
\label{sec:results_feat_importance}
We analyzed the relative importance of the different facial features for IMT-1, observing that different facial features contributed more or less, depending on the affective scale being predicted. Features' relative importance was calculated by learning the models as described above, while using only a single type of features at each time, and comparing the predictions' success rates. For example, we observed that for IMT-1 the prediction of \emph{valence} relied on all 4 feature types (including moments, discrete state features, dynamic features and miscellaneous features), while the prediction of \emph{arousal} didn't use the miscellaneous features at all, but relied heavily on the dynamic aspects of the facial expression. Similarly, the prediction of \emph{likability} utilized the miscellaneous features the most, while not using the moments features. Specifically, prediction with only miscellaneous features achieved correlation of $R=0.275$ with the \emph{likability} score, and $R=0.287$ with the \emph{rewatch} score. These observations are summarized in Figure~\ref{fig:feat_cont}. 

\begin{figure}[!ht]
  \centering
  
  \subbottom[][Valence]{\includegraphics[width=.35\textwidth]{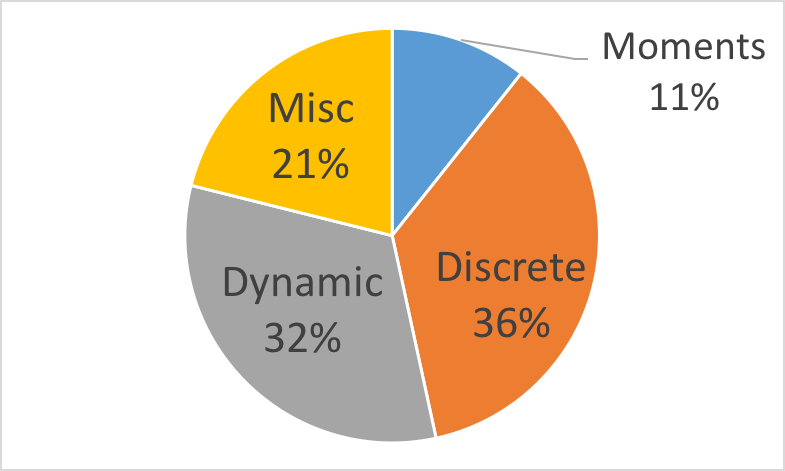}}\quad
  \subbottom[][Arousal]{\includegraphics[width=.35\textwidth]{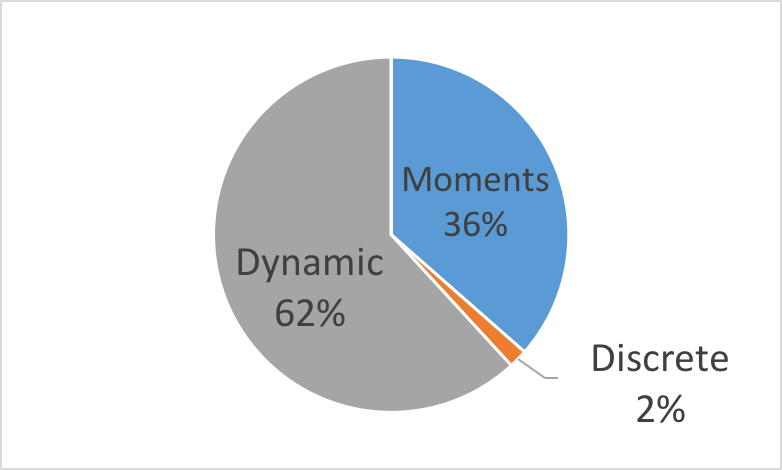}}\\
  \subbottom[][Likability]{\includegraphics[width=.35\textwidth]{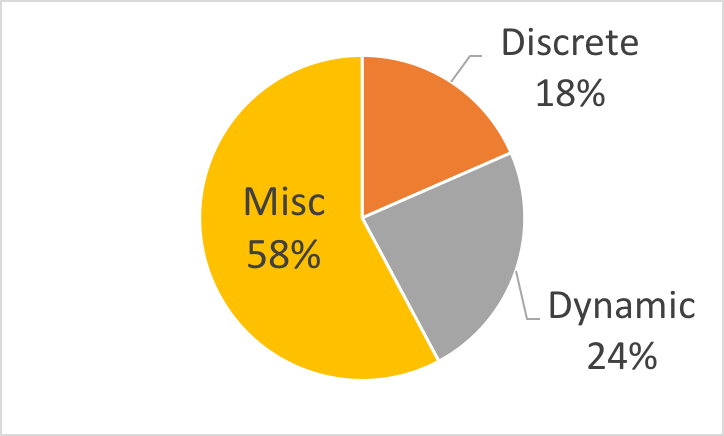}}\quad
  \subbottom[][Rewatch]{\includegraphics[width=.35\textwidth]{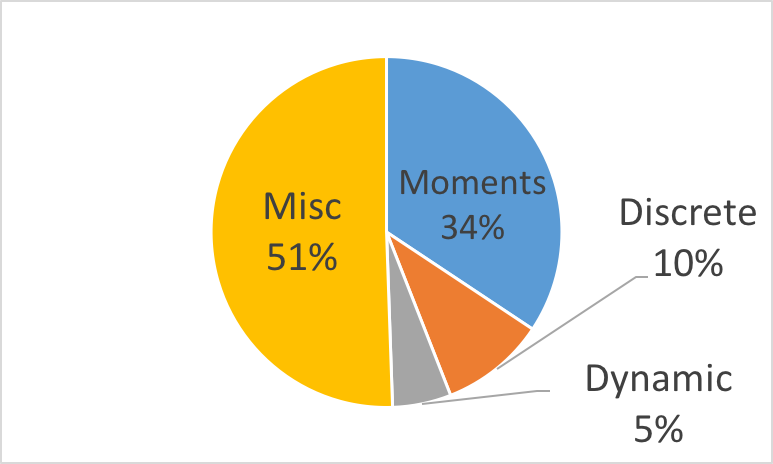}}
  
  \caption{The relative contribution of different feature groups to IMT-1.}
  \label{fig:feat_cont}
\end{figure}

As a comparison, we analyzed relative importance of the different facial features for the analogous affect prediction algorithm (namely AP-1). As can be seen in Figure~\ref{fig:feat_cont_ap}, the distribution is similar in both models, except that the dynamic features' relative importance is consistently higher in affect prediction ($\mu=+11.75\%$). Because the dynamic features are relatively more dominant in predicting subjective emotion, this observation encourages us to hypothesize that the temporal aspects of viewers' facial behavior, in addition to serving as a good predictor for emotions in general, could be used as a distinguishing property between different viewers. This idea is in line with \cite{tron2016facial}, where it was shown that these aspects helps distinguishing between schizophrenia patients and healthy individuals.

\begin{figure}[!ht]
  \centering
  
  \subbottom[][Valence]{\includegraphics[width=.35\textwidth]{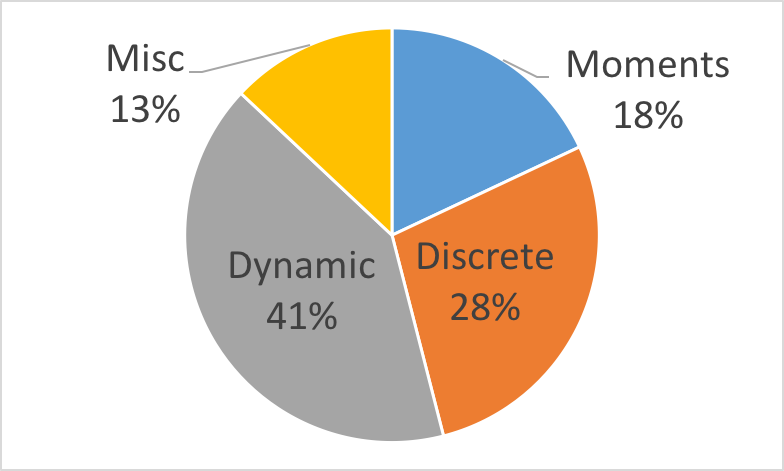}}\quad
  \subbottom[][Arousal]{\includegraphics[width=.35\textwidth]{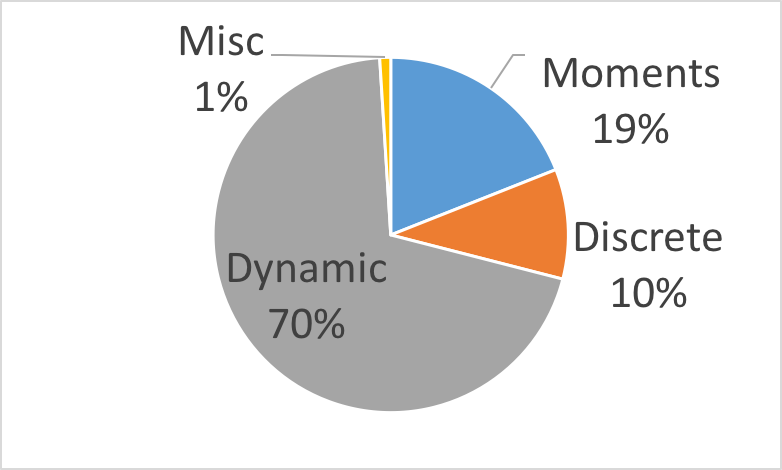}}\\
  \subbottom[][Likability]{\includegraphics[width=.35\textwidth]{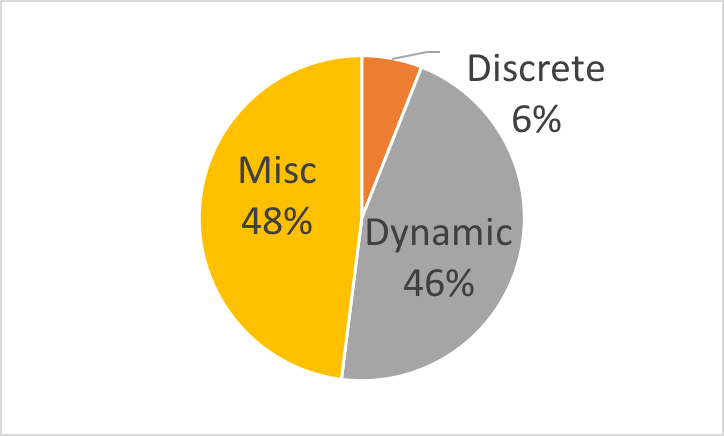}}\quad
  \subbottom[][Rewatch]{\includegraphics[width=.35\textwidth]{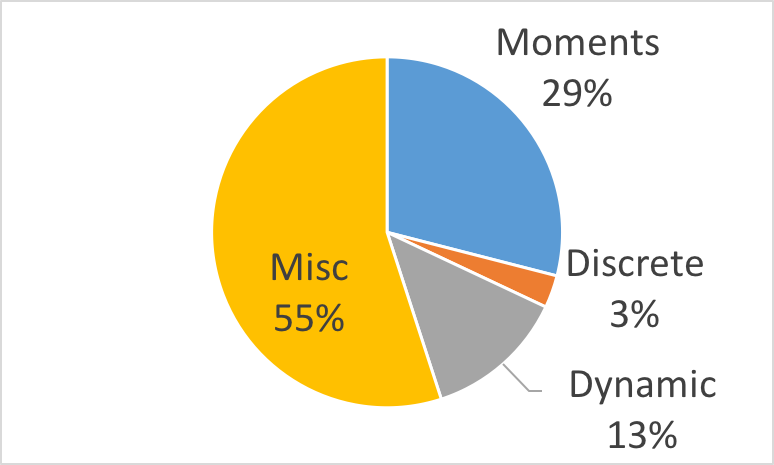}}
  
  \caption{The relative contribution of different feature groups to AP-1.}
  \label{fig:feat_cont_ap}
\end{figure}

\section{Localization of Highlight Period} 
\label{sec:results_hp}

We also analyzed the relative location of the response highlight period (HP) within the clip. Although this period was computed bottom-up from the facial recording of each individual viewer and without access to the observed video clip, the correspondence between subjects was notably high ($ICC=.941$). Not surprisingly, the beginning of the period was usually found a few seconds before the clip's end  ($\mu=-7.22$, $\sigma=4.14$), and in some clips it lasted after the clip ended (specifically in 8 out of the 18 clips). Yet the HP localization clearly depended, in a reliable manner across viewers, on the viewed clip. For example, when viewing a car safety clip, the average HP started 14 seconds before its end, probably because a highly unpleasant violent sequence of car crashes had began a second earlier. We may conclude that the HP tends to focus around the clip's end most of the times, but clip-specific analysis is preferable in order to locate it more precisely. The distribution of HPs is presented in Figure~\ref{fig:hp_local}.

\begin{figure}[!ht]
\centering
\includegraphics[width=0.7\textwidth]{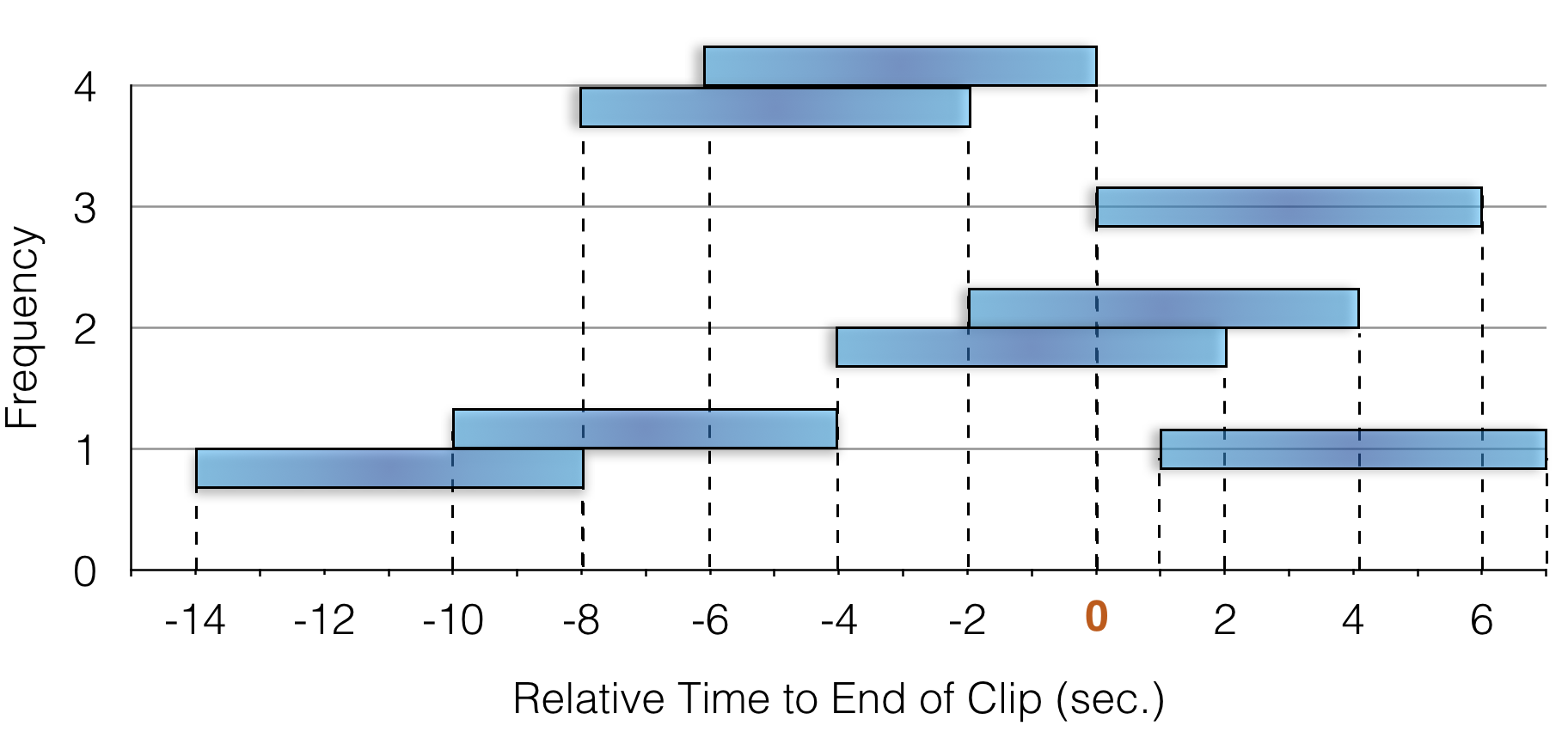} 
		\small	
		\caption{Histogram of HPs relative to the clips' end time, which marks the origin of the $X-$axis ($\mu=-7.22$, $\sigma=4.14$, $\chi^2=0.86$).
        \label{fig:hp_local}}
\end{figure}

\clearpage

\let\textcircled=\pgftextcircled
\chapter{Discussion}
\label{chap:discussion}

\initial{O}ur contribution in this work is two fold. First, we obtained a database of video clips which give rise to strong predictable emotional response, as verified by an empirical study, and which is available to the community. Second and more importantly, we described several algorithms that can predict emotional response based on spontaneous facial expressions, recorded by a depth camera. Our method provided a fairly accurate prediction for 4 scores of affective state: \emph{Valence, Arousal, Likability}, and \emph{Rewatch} (the desire to watch again). We achieved high correlation between the predicted scores and the affective tags assigned to the video. In addition, our results suggest that a group of viewers performs better as a predictor to media tagging than a single viewer (similarly to idea of "The Wisdom of Crowds"), as IMT-2 achieves evident higher success rates than IMT-1. Hence, in real-life systems, it's preferable to rely on the facial behavior of a group of known viewers than a single one, if possible.

When using facial expressions for automatic affect prediction, We saw that it's easier to predict the \emph{expected} affective state (\emph{i.e.} implicit media tagging) than the viewer's \emph{reported} affective state (\emph{i.e.} affect prediction). Further analysis evinced that a prediction based on the media tags provides a more accurate estimation for the viewers' affective state than a prediction based on the viewer's report. These results are rather surprising, and we believe that further effort to improve the affect prediction models could obtain better results. One possible course of action to expand the predictive power is utilizing viewers personality properties by using the Big-5 Personality Traits \cite{john1999big}, that were also collected in this study. Another approach is considering head and upper-body gestures, that could be also captured by depth cameras, as it has been showed that body movements contribute to emotion prediction \cite{atkinson2007evidence}.

Interestingly, when computing the period of strongest response in the viewing recordings, we saw high agreement between the different viewers. Further analysis revealed that different types of facial features are useful for the prediction of different scores of emotional state. For example, we saw that simply counting the viewer's smiles and blinks (miscellaneous features) provided an inferior, yet significantly correlated, prediction of \emph{Likability} and \emph{Rewatch}. For commercial applications, these facial features can be obtained from the laptop's embedded camera (in a similar manner  to \cite{mcduff2014automatic}). Furthermore, we found that the dynamic aspects of facial expressions contribute more to the prediction of viewers affective state than to the prediction of media tags.

In a wider perspective, we recall that on April 2005, an 18 seconds video clip titled "Me at the zoo" became the first video clip uploaded to YouTube. In the decade since, the world had witnessed an unprecedented growth of online video resources -- on YouTube alone there are over 1.2 billion video clips; adding other popular websites like Vimeo, Dailymotion and Facebook, and we reach an un-grasped amount of cuddling cats, soccer tricks and recorded DIY manuals. On April 2016 the CEO of Facebook, Mark Zuckerberg, stated that within 5 years Facebook will be almost entirely composed of videos; and with 1.8 billion active users that watch over 8 billion videos per day \cite{constine2015facebook}, that is a statement that should be taken seriously.

Such a staggering amount of videos poses many challenges to computer scientists and engineers. Since every user can upload videos as he desire, classifying and mapping them for accurate and quick retrieval, ease of use, search engines and recommendation systems becomes a challenging task. One solution is tagging the videos -- assigning descriptive labels that aids indexing and arranging them. Another challenge is to understand the viewers' expected emotional response, to refine personal customization and to comprehend the effect of the videos on the users. These are the challenges we tackled in this work.

\clearpage
%
%
\appendix
\begin{appendices}
\chapter{Review of Emotion Elicit Databases}
\label{app:app01}

\initial{T}he following pages includes a review of all major emotion eliciting databases. To be noted that the databases reviewed here might be partially to the ones released, as only the emotion eliciting clips are discussed. For example, both DEAP \cite{koelstra2012deap} and MAHNOB-HCI \cite{soleymani2012multimodala} contains major parts of subjects' physiological signals that are not mentioned in this review. Drawbacks are numbered with respect to Section~\ref{sec:db_criteria}

\begin{landscape}
    \centering
    {
    \setlength{\tabcolsep}{10pt}
\def\arraystretch{1.5}

\begin{longtable}[tb]{>{\centering}m{3cm} m{5cm} m{4cm} m{7cm}}

    \textbf{Reference} & \multicolumn{1}{c}{\textbf{Description}} & \multicolumn{1}{c}{\textbf{Affect Descriptors}} & \multicolumn{1}{c}{\textbf{Drawbacks}} \\
    \hline \hline
    
    \textbf{LIRIS-ACCEDE} \cite{baveye2015liris} & 9,800 excerpts extracted from 160 feature films and short films, shared under Creative Commons license. Duration of 8-12 seconds, rated online by over 2,000 crowdsourced participants. & Valance and Arousal on 5-point scale (derived from rating-by-comparison). & (1) Relatively short; (6) Familiarity wasn't examined; (7) Crowdsourced. \\ \hline
    
    \textbf{VideoEmotions} \cite{jiang2014predicting} & 1,101 YouTube (480) and Flickr (621) clips that were returned as search results for 8 different emotional adjectives (\emph{e.g.} Anger, Joy and Disgust). Average duration of 107 seconds. & Search results labels: Anger, Anticipation, disgust, Fear,
Joy, Sadness, Surprise, and Trust. & (1) 107 seconds in average; (2) No V-A ratings; (6) As above. \\ \hline

    \textbf{EMDB} \cite{carvalho2012emotional} & 52 non-auditory 40-second excerpts from commercial films, rated by 113 subjects. & Valence, Arousal and Dominance on discrete 9-point scales, and free language text concerning felt emotion during and after watching. & (4) The lack of sound may be a confounding factor (because it's unusual) and alters the subjects' viewing experience; (6) Taken from well-known (partially Oscar winning) movies; (8) Most films are copyright protected. \\ \hline
    
    \textbf{DEAP} \cite{koelstra2012deap} & 120 1-minute excerpts (with maximum emotional content) from music video clips, rated by 14-16 subjects per clip. & Valence, Arousal and Dominance on discrete 9-point scales. & (1) Relatively long; (4) All clips are music videos. Although they differ in their emotional tags, they have many similarities (\emph{e.g. } they are all accompanied by songs and dancing); (6) Many of them are clips of highly popular songs; (7) Rated by online subjects; (8) Some of the clips are not available online anymore due to copyright claims. \\ \hline

    \textbf{MAHNOB-HCI} \cite{soleymani2012multimodala} & 20 excerpts from Hollywood movies, YouTube clips and weather reports. Duration of 35-117 seconds ($\mu=81.4$), rated by 27 participants. & Valence, Arousal, Dominance and Predictability on discrete 9-point scales, as well as basic emotion tagging. & (1) Relatively long; (3) Major sensor presence (EEG recordings, videotaping, audio recordings, physiological signals recording and eye tracking); (6) Most clips are from Hollywood famous movies; (8) Most films are copyright protected. \\ \hline
    
    \textbf{FilmStim} \cite{schaefer2010assessing} & 70 excerpts from dozens of films selected by 50 film rental store managers. Duration of 1-7 minutes long, dubbed (or originally) in French, rated by a total of 364 subjects. & Arousal level and 16 emotional adjectives on a 7-point scales (\emph{e.g. } Anxious, Tense, Nervous), Positive and Negative Affect Schedule (PANAS scale \cite{watson1988development}). & (1) Over a minute long; (5) French Speaking; (6) Taken from well-known (partially Oscar winning) movies; (8) As above. \\ \hline
    
    \textbf{Untitled} \cite{wang2006affective} & 36 Full-length Hollywood movies (a total of 2040 scenes), rated by 3 subjects. & Most suitable emotion for each scene, out of 7 basic emotions. & (1) Assuming the average movie length is 1 hour and 30 minutes, the average scene length is about 95 seconds; (2) No V-A Rating; (6) As above; (8) As above. \\ \hline
    
    \textbf{Untitled} \cite{gross1995emotion} & 78 excerpts extracted from over 250 films, 8-1192 seconds ($\mu=151$), rated by at least 25 subjects per clip, with 494 subjects in total. & 16 basic emotions on a 9-point scale. & (1) Average of 2:31; (2) As above; (6) As above; (8) As above. \\ \hline
    
    \textbf{Untitled} \cite{philippot1993inducing} & 12 excerpts extracted from 9 films. Duration of 3-6 minutes, dubbed (or originally) in French, rated by 60 subjects. & 10 basic emotions on a 5-point scale, nine scales of semantic feeling (\emph{e.g. } "Little -- Large") and free labels. & (1) Relatively long; (5) French Speaking; (6) As above; (8) As above. \\ \hline

\caption{Emotion eliciting video clips databases.}
\end{longtable}

    }
\end{landscape}\end{appendices}
\clearpage
%
\backmatter

\bibliographystyle{siam}
\refstepcounter{chapter}
\bibliography{main}
%
%
\end{document}